\documentclass[onecolumn, superscriptaddress, prmaterials, amsmath,amssymb,noeprint,12pt]{revtex4-2}

\usepackage{hyperref}
\hypersetup{backref,colorlinks=true,allcolors=prussianblue}

\usepackage{graphicx}
\usepackage[caption=false]{subfig}

\usepackage[table]{xcolor}
\definecolor{prussianblue}{rgb}{0.0, 0.19, 0.33}

\usepackage{setspace}
\usepackage{float}
\usepackage{indentfirst}

\usepackage{tcolorbox}

\usepackage{miller}

\def\eq#1{Eq.~\eqref{eq:#1}}
\def\fig#1{Fig.~\ref{fig:#1}}
\def\tab#1{Table~\ref{tab:#1}}

\def\si{\textbf{SI}}

\UseRawInputEncoding

\begin{document}

\author{Musanna Galib}
\author{Okan K. Orhan}
\author{Mauricio Ponga}
\address{Department of Mechanical Engineering, University of British Columbia, 2054 - 6250 Applied Science Lane, Vancouver, BC, V6T 1Z4, Canada}

\title{Engineering Chemo-Mechanical Properties of Zn Surfaces via Alucone Coating}

\begin{abstract}

Aqueous zinc (Zn)-ion batteries (AZIB) are promising candidates for the next-generation energy store systems due to their high capacity and low cost. 
Despite their nominal performance, Zn anodes tend to rapidly develop dendrite and fracture, leading to substantial capacity loss and cycling stability failure. 
Well-controlled coating using organic-inorganic hybrid molecules is highly promising to substantially improve their chemo-mechanical stability without compromising their performance. 
We herein present a critical assessment of the chemical and mechanical stability of alucone-coated Zn surfaces using first-principles simulations.
Negative adsorption energies indicate strong cohesive strengths between alucone and the selected Zn surfaces. 
Energetically favorable alucone coatings are further verified by charge transfer at interfaces as seen through Bader charge analysis.
Negative surface stress profiles at alucone coated interface are mostly responsible for surface reconstruction.
The contributions of surface elastic constants are dependent on the selection of slip planes and the thickness of the thin film.
By considering plane stress conditions, we calculate the mechanical properties  which indicate the ductility of the alucone-coated basal thin film. 
\end{abstract}

\maketitle

\section{Introduction}
Fossil-fuel consumption in transportation technologies is the second largest source of $\mathrm{CO_{2}}$ emission, and expected to increase by 28\% by 2040~\cite{10.1016/j.apenergy.2019.113404}.
Greener energy applications highly depends on the development of well-performing, reliable, and cost-effective energy storage systems such as batteries, fuel cell, and supercapacitors~\cite{Winter_Brodd,Larcher_Tarascon, 10.1007/s00894-020-04483-5}.
Regardless of any battery design, electrodes are the main components that determine the bottleneck of battery capacity, power output, cyclic stability, and lifetimes of energy storage systems~\cite{Bilal_Chuan_Husam, 10.1021/acsenergylett.8b01552}.
Thus, fabrication and surface engineering of electrode materials are the most vital aspect of the next-generation energy storage systems~\cite{Bilal_Chuan_Husam}. 

Aqueous zinc-ion batteries (AZIBs) have attracted considerable interest due to their exceptional storage capacity of $\sim 5854$~mAh$\cdot$cm$^{-1}$ ($\sim 820$~mAh$\cdot$g$^{-1}$) compared to conventional Li batteries with a storage capacity of $\sim 2042$~mAh$\cdot$cm$^{-1}$~\cite{10.1039/C8TA12014E}.
AZIBs generally consist of a Zn anode and  an (in)organic cathodes such as MnO$_{2}$, an aqueous electrolyte (mildly acidic pH) such as ZnSO$_{4}$, and a separator such as a ceramic separator\cite{10.1016/j.jpowsour.2018.05.072}.
Zn is non-toxic and abundant in Earth's crust~\cite{10.1021/acsenergylett.8b01552}.
It can be oxidized to Zn$^{2+}$ during stripping (discharging) without forming any intermediate phases~\cite{10.1002/adfm.201802564} under mildly acidic conditions (pH = $4-6$) which promote the reversible Zn stripping/plating (discharging/charging) process~\cite{10.1039/D0SC00022A,10.1016/j.mtener.2017.12.012, 10.1002/anie.201106307}.
It also has a redox potential is $-0.76$~V for Zn, for which  the hydrogen-evolution reaction provides an over-potential suitable for effective battery operation~\cite{10.1039/D0SC00022A}.
Thus, AZIBs are highly favorable for flexible and wearable batteries  as they are safe, sustainability, cost-effective, environmentally friendly, and provide reliable power output~\cite{10.1021/acsnano.1c01389, 10.1039/C7EE03232C}.

Despite their highly promising premise, AZIBs suffer two major drawbacks, limiting their widespread commercial  applications\cite{10.1021/acsami.0c06009}. 
The first drawback is stability issues in $\mathrm{MnO_{2}}$ cathodes. 
The second one is  dendrite formation on the Zn anode surface due to irregular plating/stripping of Zn-ions during the charging/discharging cycles.  
This commonly leads to a short circuit between the cathode and anode~\cite{10.1149/1.1838919}. 
A suggested strategy to overcome rapid dendrite formation in Zn anodes is nano-structured coating~\cite{10.1016/j.mtener.2017.12.012}. 
For this purpose, atomic layer deposition (ALD)~\cite{Johnson_Hultqvist} and molecular layer deposition (MLD)~\cite{Sundberg_Karppinen} are the state-of-art techniques to achieve well-controlled thickness and structure~\cite{Zhao_Zhang_Liu}, superior conformity~\cite{Indrek_Martti_Pars, Basuvalingam_Bloodgood_Verheijen}, and uniformity~\cite{Elers_Blomberg_Peussa}. 
It has been shown that ALD and MLD are more reliable for surface engineering compared to more conventional coating techniques such as physical vapor deposition (PVD) or chemical vapor deposition (CVD)~\cite{Zhao_Zhang_Liu, 10.1080/14686996.2019.1599694}.
Furthermore, ALD/MLD allow tuning compositions easily where the deposition temperatures are relatively low ~\cite{Zhao_Zhang_Liu}. 
Thus, they became quite popular in the field of surface engineering for energy storage systems, which require high-level accuracy and tunability in surface construction~\cite{Leskela_Ritala_2003,Leskela_Ritala_2002}.
While ALD is mostly limited to inorganic coating agents, MLD provides more diverse coating agents such as hybrid organic-inorganic coating at the expense of a more complex coating process~\cite{Zhao_Zhang_Liu}.  
However, MLD opens up virtually limitless possibilities in terms of coating agents, enabling superior control over thermodynamic and mechanical properties at interfaces~\cite{Goncharova}. 

There have been previous works on the nano-structured coating of Zn anodes such as the modified polyamide coating~\cite{10.1039/C9EE00596J}, the ultra-thin TiO$_2$ coating using ALD~\cite{ 10.1002/aenm.201801090} and the drop-casting of nanoporous CaCO$_3$ \cite{10.1002/admi.201800848}. 
Despite improving mechanical stability, these ceramic and polymeric coatings severely hinder the performance of Zn anodes by significantly increasing surface resistance, leading to poor Zn ions diffusion~\cite{10.1021/acsami.0c06009}. 
With that in mind, an effective coating agent is expected to satisfy the four major conditions: (i) restricting free diffusion of Zn ions by electrostatic interaction or by physical restriction,  (ii) facilitating ionic conduction only and restricting electronic conduction, (iii) keeping the Zn electrode stable in an electrolyte, (iv) maintaining sufficient toughness and rigidity to adjust to the volume change throughout the charging/discharging process. 
Thus, it is quite challenging to select and deposit an effective and cost-efficient coating material that suppresses the dendrite formation while maintaining battery performance \cite{10.1021/acsami.0c06009}.

Aluminum alkoxide (alucone)~\cite{10.1021/ja00265a024} is a promising candidate which has been already used stabilize sulfur (S) cathodes and silicon (Si) anodes in Li--S\cite{10.1039/C4CC04097J} and Li--ion batteries~\cite{10.1002/adma.201304714}, respectively.
It is produced by chemical reactions of organic alcohols and trimethylaluminum (TMA) precursors during the MLD process.
Alucone possibly suppresses dendrite formation while not hindering charge transfer in charging/discharging cycles by providing  flexible gas-diffusion barriers, superior corrosion resistance, and wettability for Zn anodes~\cite{10.1063/1.4766731, 10.1039/D0TA07232J}.
Thus, it is crucial to understand the effects of alucone coating on the thermodynamic, chemical, and mechanical properties of Zn anodes for effective and safe AZIB applications. 
However, it is a challenging task to experimentally study underlying mechanisms as well as limited only to a small number of coating agents and samples.
First-principles simulations offer powerful and feasible tools to study surface chemistry and properties of complex structures~\cite{10.1016/S0039-6028(99)00741-4, 10.1103/PhysRevB.74.165402}. 
Tran \emph{et al.} have calculated surface energies of 70 elements using high-throughput first-principles approaches~\cite{10.1038/sdata.2016.80}. 
Furthermore, they are also highly accurate to capture charge transfer at the interface~\cite{10.1039/C4CP02741H}.
For instance, Lawson \emph{et al.} have studied the interaction between Al$_2$O$_3$, HfO$_2$ and MgO surfaces and MoS$_2$ coating during the first cycle of ALD.

In this work, we present an accurate first-principles investigation of thermal and chemical stability and mechanical properties of alucone-coated Zn surfaces. 
Crystal and electronic structures are optimized for the \hkl(0 0 0 1) and \hkl(1 -1 0 0) surfaces of the hexagonal closed-packed (HCP) Zn with and without single-layer alucone to calculate adsorption energies and analyze charge transfer at the interfaces.
It is shown that the alucone-coated surfaces are energetically and chemically stable.
Using the pristine bulk Zn, its surfaces, and thin films, we approximate the elastic properties at the interfaces of the alucone-coated surfaces.
The mechanical properties analysis indicates the ductility of alucone-coated basal thin film however these properties are direction-dependent.
Finally, we provide crucial figures-of-merits (FOMs) for substrate + coating selection based on their individual electric properties.

\section{Theoretical and Computational Methodology}
Kohn-Sham density-functional theory (KS-DFT)~\cite{10.1103/PhysRev.136.B864,10.1103/PhysRev.140.A1133} using semi-local exchange-correlation functionals~\cite{10.1103/PhysRev.140.A1133} provides  feasible and accurate tools for studying ground-sate properties of solids. 
Beyond structural and electronic properties, the approximate KS-DFT is also suitable to expediently calculate elastic properties using   Hooke's law.
In this work, its plane-wave-based implementation within the Vienna Ab-Initio Simulation Package (VASP)~\cite{PhysRevB.47.558,KRESSE199615,PhysRevB.54.11169} using the projector augmented-wave (PAW) pseudo-potentials~\cite{PhysRevB.59.1758}. 
The exchange-correlation functional was expressed by the generalized gradient approximation (GGA) of Perdew, Burke, and Ernzerhof (PBE) ~\cite{Perdew_Burke_Ernzerhof}.    
The total energy is minimized in the DFT calculation for the occupation of Kohn-Sham's self-consistent scheme.
In this study, partial wave occupancies were computed by Gaussian smearing, and the width of smearing was \(\sigma\)=0.026 eV  which corresponds to 300~K.
Electronic  energies  were  computed  with the  tolerance of $\mathrm{10^{-6}}$ eV  using  a  self-consistent-field (SCF).
\par One of the key features of a crystal/surface is to determine a relaxed geometry, that is to find the geometry with the lowest energy using the geometry relaxation test.
As a result, the unit cell of the crystal goes through relaxation in all degrees of freedom (ionic positions, cell shape, and volume) to find the lowest energy model.
The conventional convergence criterion for kinetic-energy cutoff $\mathrm{{\it{E}}_{cut}}$ is to attain a change of less than 1 meV/atom in total energy for raising the $\mathrm{{\it{E}}_{cut}}$. 
The PAW pseudopotential was exercised to report the core electrons whereas the valence orbitals were illustrated using a cutoff kinetic energy of 360 eV with plane-wave basis set as it fulfills the convergence criterion.

The Zn unit cell was optimized with a 15 x 15 x 15 k-points mesh to formulate the Zn surface after the convergence test, such as the entropy< 1 meV per atom.
A Zn \hkl(0001) surface with six Zn layers was constructed using the optimized Zn unit cell (Supporting Information (\si)~\fig{no_of_free_layer_optimization}).
A 15 x 15 x 1 Monkhorst--Pack k-points mesh for the Zn surface model was employed to produce the plane wave basis set.
\begin{figure}[h]
\centering
\includegraphics[height=2in]{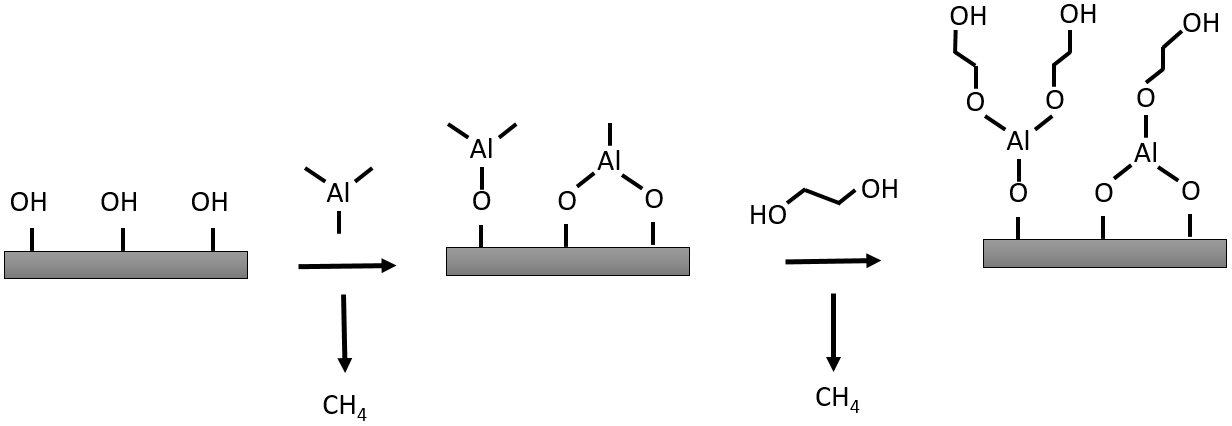}
\caption{Schematic illustration of the molecular layer deposition of alucone on a substrate.\label{fig:Alucone-MLD}}
\end{figure}
In~\fig{Alucone-MLD}, the MLD of a single-layer alucone on a substrate is schematically shown.
It is a two-step process for depositing single layer of alucone~\cite{Dameron_Seghete_Burton}. 
Alucone is the general name of polymeric aluminum alkoxide compounds which have a general sequence of as $\mathrm{\cdots Al-O-R-O-Al \cdots}$.
The radical R is an organic molecule. 
Alucone is deposited on a substrate by first pulsing TMA  on to the hydroxylated substrate,  purging CH$_4$ as by-product. 
At the second stage, the activated surface reacts with ethylene glycol (EG) while purging additional CH$_4$. 
These two step deposition process can be chemically expressed as~\cite{Liang_Weimer, Dameron_Seghete_Burton}
\begin{align}\label{chem:ch1}
\mathrm{-OH^{*} + Al(CH_{3})_{3}} &= \mathrm{-OAl(CH_{3})^{*}_{2}+CH_{4}}\nonumber \\
\mathrm{-AlCH^{*}_{3} + OHCH_{2}CH_{2}OH} &= \mathrm{ AlOCH_{2}CH_{2}OH^{*}+CH_{4}},
\end{align}
where the asterisks indicate the surface species.
By repeating this two-step process, the thickness of alucone coating can be linearly increased. 
In this work, we only focus on the Zn surfaces which are fully coated with a single layer of alucone.
\subsection{Van der Waals Interactions}
London dispersion forces, generally mentioned as van der Waals (vdW) forces, develops a weak instantaneous
interaction that varies as $\sim1/r^{6}$ decay~\cite{10.1007/BF01341258}.
These interactions form due to the charge fluctuation between dipoles. 
However, the contribution of vdW interactions is absent in standard DFT exchange-correlation (XC) functionals. 
This limitation puts an error bar on the calculation of the adsorption and cohesive energy of weakly bonded molecules on surfaces. 
In this study, we used DFT-D2 semi-empirical correction method suggested by Grimme~\cite{10.1002/jcc.20495}, which adds a correction term ($\mathrm{{\it{E}}_{disp}}$) to the total energy after
every self-consistent cycle\cite{10.1088/2053-1591/3/4/046501}.
\begin{align}\label{vdw_DFT_D2}
E &= E_{\textrm{KS}}+E_{\textrm{disp}} \\
E_{\textrm{disp}} &= -S_{\textrm{6}}\sum_{i}^{N-1} \sum_{j=i+1}^{N}\frac{C_{6}^{ij}}{(R^{ij})^{6}} F_{\textrm{damp}}(R^{ij})\\
F_{\textrm{damp}}(R^{ij})&=\frac{1}{1+e^{d(\frac{R^{ij}}{R_{0}^{i}+R_{0}^{j}-1})}}\\
C_{\textrm{6}}^{ij} &= \sqrt{C_{\textrm{6}}^{i}C_{\textrm{6}}^{j}}
\end{align}
where ${E}_{\textrm{disp}}$ is dispersion term, $F_{\textrm{damp}}$ is damping function, $ C_{6}^{ij}$ is dispersion coefficient, $S_{6}$ is the global scaling factor, $R_{0}^{i}$ is the vdW radius of atom i, $R^{ij}$ is the distance between atom number i and j, $C_{6}^{i}$ is the atomic parameter, and d is a damping parameter.
\subsection{Energetic and Chemical Stability}
Surface energy ($\gamma$) is a central quantity due to its analytical relation to adsorption energy and surface elastic properties. 
It is defined as the energy required to separate a bulk solid into two pieces. 
Despite its simple definition, it is quite challenging to obtain $\gamma$ within experimental methods. 
For instance, it generally requires estimating surface tension for metals at their melting temperatures~\cite{Wang_Jian, 10.1016/S0039-6028(98)00363-X}.
Thus, it is generally calculated using first-principles simulations. 
Conventionally, it is given by~\cite{Sun}
\begin{align} \label{eq:standard_surface_energy}
\gamma = \frac{1}{2A} \Big[E_\textrm{slab}-N E_\textrm{bulk}\Big],
\end{align} 
where $E_\textrm{slab}$ is the ground-state energy of the slab with the surface area of $A$ and consisting of $N$ atoms, and $E_\textrm{bulk}$ is the ground-state energy of the bulk per atom. 
In practice, $E_\textrm{slab}$ and $E_\textrm{bulk}$ are obtained through separate simulations which may lead to some convergence problems~\cite{Sun, Scholz}. 
Thus, there are some modified expressions available~\cite{Scholz}.
In this work, \eq{standard_surface_energy} works without any convergence problem and with similar accuracy with more complex approaches (see \si~ for further details and comparison of various approaches). 

\begin{figure}[h]
\vspace{3mm}
\begin{center}
\includegraphics[width=9cm]{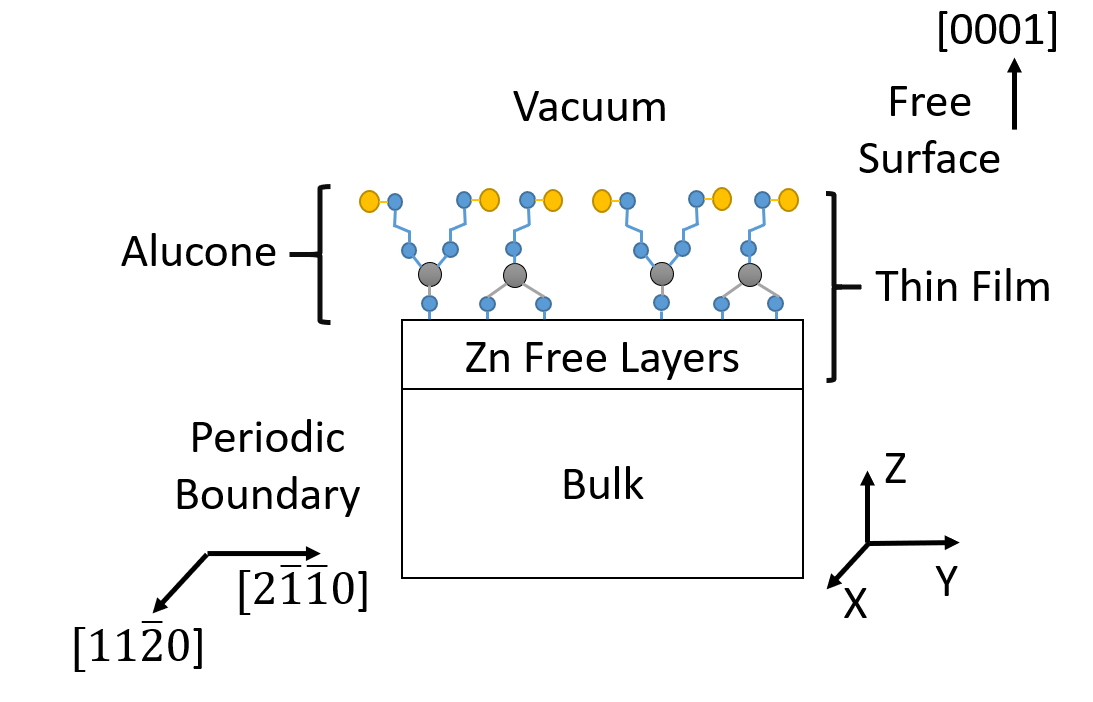}
\caption {Thin film model.\label{fig:surface_model}}
\end{center}
\vspace{-8mm}
\end{figure}

While the surface energy is crucial to assess the energetic stability of a pristine surface, the adsorption energy ($E_\mathrm{ads}$) is used to assess whether an absorbate (molecule) is likely to be chemically attached to an adsorbent (surface), given by ~\cite{10.1039/C6RA08958E, 10.1103/PhysRevB.68.245409}:
\begin{align}\label{eq:eq2}
E_\mathrm{ads} = E_\mathrm{surf+mol} - \left(E_\mathrm{surf}+E_\mathrm{mol}\right),
\end{align}
where $E_\mathrm{surf+mol}$ and $E_\mathrm{surf}$ are the total energy of the slab with and without the absorbate, respectively. 
$E_\mathrm{mol}$ is the total energy of the absorbate in the vacuum. 
In practice, a surface has to be constructed as a thick slab with large vacuums on both sides as a semi-infinite structure is not defined in plane-wave-based implementations of the approximate KS-DFT due to mismatching boundary conditions~\cite{Sun}.  
We defined the model with a periodic and free boundary conditions for \hkl[11-20], \hkl[2-1-10] and  \hkl[0001] directions, respectively, which is shown in~\fig{surface_model}.
For prismatic systems, periodic boundary conditions are in (1/3)\hkl[11-20] and \hkl[0001] direction and free boundary in \hkl[1-100] direction.
In the case of full-coverage coating,~\eq{eq2} can be expressed either per absorbate or per unit surface area.

\eq{eq2} serves as a FOM for the energetic stability of a coated surface. 
Addition to that, the charge transfer between them indicates the chemical stability due to chemical bonding. 
Although, the charge transfer is far more complex, the charge-density difference given by~\cite{Lawson}
\begin{align}\label{eq:eq3}
\Delta\rho = \rho_\mathrm{surf+mol} - \left(\rho_\mathrm{surf}+\rho_\mathrm{mol}\right),
\end{align} 
where  $\rho_\mathrm{surf+mol}$ and $\rho_\mathrm{surf}$ are the charge densities of the coated and pristine surfaces, respectively, and  $\rho_\mathrm{mol}$ is the charge density of the molecular thin film.
The three charge densities are required to be computed on appropriately matching and relative position at a common coordinate system.

\subsection{Surface Mechanical Properties}
The surface-stress profile is crucial to assess the mechanical stability of a surface. 
The strain-dependent surface stress is commonly given by~\cite{10.1103/PhysRevB.71.094104}
\begin{align}\label{eq:eq4}
\tau(\epsilon) = \gamma+ \frac{\partial \gamma}{\partial \epsilon},
\end{align}
where $\epsilon$ is the infinitesimal strain. 
On the other hand, the total energy of the deformed surface with an area $A$ due to the infinitesimal strain can be defined as
\begin{align}\label{eq:eq5}
E^\textrm{S}=A\gamma(\epsilon).
\end{align}
At the zero-strain condition ($\tau(0)=\tau^0$), the components of the surface stress tensor given in \eq{eq4} can be expressed as
\begin{align}\label{eq:eq6}
\tau_{ij}^{{\textrm{0}}} = \left[\frac{1}{A}\frac{\partial E^{{\textrm{S}}}}{\partial \epsilon_{ij}}\right]_{\epsilon=0}
\end{align}
for a small infinitesimal strain of $\epsilon$. 
Usually, we refer to stress at zero strain condition ($\tau^{0}_{ij}$) as \emph{surface stress}.
Using \eq{eq6}, the fourth-order surface elastic tensor is defined as~\cite{10.1103/PhysRevB.71.094104}
\begin{align}\label{eq:eq7}
C^\textrm{S}_{ijkl} & = \frac{\partial \tau_{ij}}{\partial \epsilon_{kl}}\Bigg|_{\epsilon=0}\nonumber \\
& =\left[2\gamma\delta_{ik}\delta_{jl}+\delta_{ij}\frac{\partial \gamma}{\partial \epsilon_{kl}}+\frac{\partial^{2} \gamma}{\partial \epsilon_{ij} \partial \epsilon_{kl}}\right]_{\epsilon=0}.
\end{align}
In practice, $\gamma$ is calculated for a small set of $\epsilon$, and it requires accurate interpolation to compute the partial derivations. 
We used three interpolation techniques, namely the finite-difference method (FDM)~\cite{10.1016/S0377-0427(00)00507-0}, the local-maximum entropy(LME)~\cite{10.1002/nme.1534} and the higher-order LME scheme (HOLMES)~\cite{10.1007/978-3-642-32979-1_7} to ensure accuracy and compare their respective performance (see \si~for further details). 

\par The surface elastic constants can be notably dissimilar from their corresponding bulk values~\cite{Liu_2014} and the elastic response of nano-structures relies on it. 
Furthermore, it is difficult to compute the surface elastic constants experimentally~\cite{10.1103/PhysRevB.71.094104, 10.1103/PhysRevB.49.10699}.
Streitz \emph{et al.} first mentioned the impact of surfaces on the multilayers and biaxial modulus of thin films using a scaling relation but did not fully formulate the surface elastic constants~\cite{10.1103/PhysRevB.49.10699}.
To illustrate the modification in elastic constants because of the exposed surface (thin-film approximation (TFA)), surface elastic constants are added with second-order bulk elastic constants considering the plane-stress condition~\cite{Hossein}.

Using Hooke's law for orthotropic materials, the plane-stress condition can be expressed as~\cite{Hossein}
\begin{align}\label{eq:eq9}
\begin{pmatrix} \sigma_{1}\\ \sigma_{2}\\ \sigma_{3}\end{pmatrix}=
\begin{pmatrix}
C_{11}^{\textrm{plane}} & C_{12}^{\textrm{plane}} & C_{13}^{\textrm{plane}}\\ 
C_{21}^{\textrm{plane}} & C_{22}^{\textrm{plane}} & C_{23}^{\textrm{plane}}\\
C_{31}^{\textrm{plane}} & C_{32}^{\textrm{plane}} & C_{33}^{\textrm{plane}}
\end{pmatrix} 
\begin{pmatrix} \epsilon_{1}\\ \epsilon_{2}\\ \epsilon_{3} \end{pmatrix},
\end{align}
which provide the relations between the  plane stresses and the bulk elastic constants for an orthotropic elastic body, given by~\cite{Hossein, LEMANT_Pineau1,Izumi}
\begin{gather}\label{eq:eq10}
C_{11}^{\textrm{plane}} = (C_{11}-\frac{C_{13}^2}{C_{33}}) \nonumber \\
C_{22}^{\textrm{plane}} = (C_{22}-\frac{C_{23}^2}{C_{33}}) \nonumber\\
C_{12}^{\textrm{plane}} = C_{21}^{\textrm{plane}} = (C_{12}-\frac{C_{12}C_{13}}{C_{33}}) \nonumber\\
C_{33}^{\textrm{plane}} = C_{66} \nonumber\\
C_{13}^{\textrm{plane}} = C_{23}^{\textrm{plane}} = C_{31}^{\textrm{plane}} = C_{32}^{\textrm{plane}} = 0,
\end{gather}
where $C_{ij}$ are components of the elastic tensor of the bulk crystals using Voigt's notation.  
Finally,  the second-order effective elastic constants can be given by~\cite{10.5772/104209, Hossein}
\begin{align}\label{eq:eq11}
C_{ij}^{\textrm{film}} = C_{ij}^{\textrm{plane}}+ \frac{2}{L_{z}}C^{\textrm{S}}_{ij},
\end{align}
where $L_z$ is the film thickness.
\subsubsection{Figures of Merits for Mechanical Performance}
Elastic constants are microscopic quantities that are well-defined. Other technologically relevant mechanical quantities can serve as a FOM to assess the quality of the coated surface.
It is particularly crucial to determine the hardness, and fracture toughness of nanostructures used in devices such as batteries to ensure safety and durability.
The Vickers hardness ($H_\mathrm{V}$) is the industry-standard FOM to assess wear resistance. 
There is no robust theoretical model as its measurement is particularly dependent on morphology, impurity of samples, choice of indenter, and direction of acting forces relative to crystallographic orientation~\cite{10.1016/j.intermet.2011.03.026}.
However, there are some semi-empirical approximations such as ~\cite{10.1557/S0883769400031420, 10.1016/j.ijrmhm.2012.02.021}
\begin{align}\label{eq:eq1}
H_\mathrm{V}^\mathrm{Teter}=0.151\;G, \quad \mathrm{and} \quad  H_\mathrm{V}^\mathrm{Tian} = 0.92(G/B)^{1.137}G^{0.708},
\end{align}
where $B$ and $G$ are the bulk and shear moduli, respectively. 
While the former tends to overestimate, the latter tends to underestimate.
Thus, we used their arithmetic averaging such as $H_\mathrm{V} = 0.5 (H_\mathrm{V}^\mathrm{Teter} + H_\mathrm{V}^\mathrm{Tian}$).

Another crucial FOM is the fracture toughness ($K_\mathrm{IC}$), which restrains the crack propagating~\cite{10.1063/5.0047139, 10.1016/j.tsf.2007.02.106, 10.1063/1.5066311}. 
Niu \emph{et al.} has proposed the semi-empirical formula, given by~\cite{10.1063/1.5066311, 10.1063/1.5113622}
\begin{align} \label{eq:eq13}
K_{\textrm{IC}} = V_{0}^\frac{1}{6}G\left(\frac{B}{G}\right)^\frac{1}{2},
\end{align} 
where $V_0$ is the volume of the unstrained structure. 

The development of stress in the thin-film deposition process is one of the least explored fundamental aspects of managing the dendrite growth problem~\cite{10.1038/s41560-018-0104-5}.
Micro-structural evolution phenomena in materials such as thin film deposition are stress-driven.
Therefore, it is important to check whether notable stress exists during alucone deposition and if it is the source of Zn dendrite growth or surface wrinkling. 
The minimum membrane strain for the beginning of wrinkling is specified by~\cite{10.1038/s41560-018-0104-5}
\begin{align}\label{eq:eq14} 
\epsilon_\textrm{m} = -\frac{1}{4}\left[\frac{3E_\textrm{sub}(1-\nu_\textrm{tf}^2)}{E_\textrm{tf}(1-\nu_\textrm{sub}^2)}\right]^\frac{2}{3}
\end{align} 
where $E_\mathrm{sub}$ and  $E_\mathrm{tf}$ are the elastic moduli of the substrate and thin film, respectively, and $\nu_\mathrm{sub}$ and  $\nu_\mathrm{tf}$ are Poisson's ratios of the substrate and thin film.
\section{Results and Discussion}

Second-order elastic properties of Bulk Zn have been computed and verified with experimental data as shown in \tab{zinc_crystal_elastic_constants}. 
For Zn thin film,  we choose a model of four fixed layers and two free layers which have been justified by calculating surface layer relaxation as shown in \fig{no_of_free_layer_optimization} and \fig{no_of_fixed_layer_optimization}.
\subsection{Structural Optimization of Alucone-Coated Zn Surface}
The Al-O bond length varies from 1.7 - 1.9~\r{A} depending on the bonding configuration.
The C-O, C-H, and C-C bond lengths are approximately 1.1~\r{A}, 1.1~\r{A}, and 1.4~\r{A}, respectively.
Through a series of extensive convergence tests (5, 10, 20, 30~\r{A}), it was verified that 10~\r{A} vacuum were adequate to confirm that the surface energies were converged with variation less than 1 $\mathrm{meV/\mathring{A}^2}$ along with avoiding inter-slab interaction.
The Van der Waals energy was approximately 5 percent with respect to the total ground state energy (\fig{vdw}). 

\begin{figure}[t]
\centering
\subfloat[]{
\label{}
\includegraphics[height=2.70in]{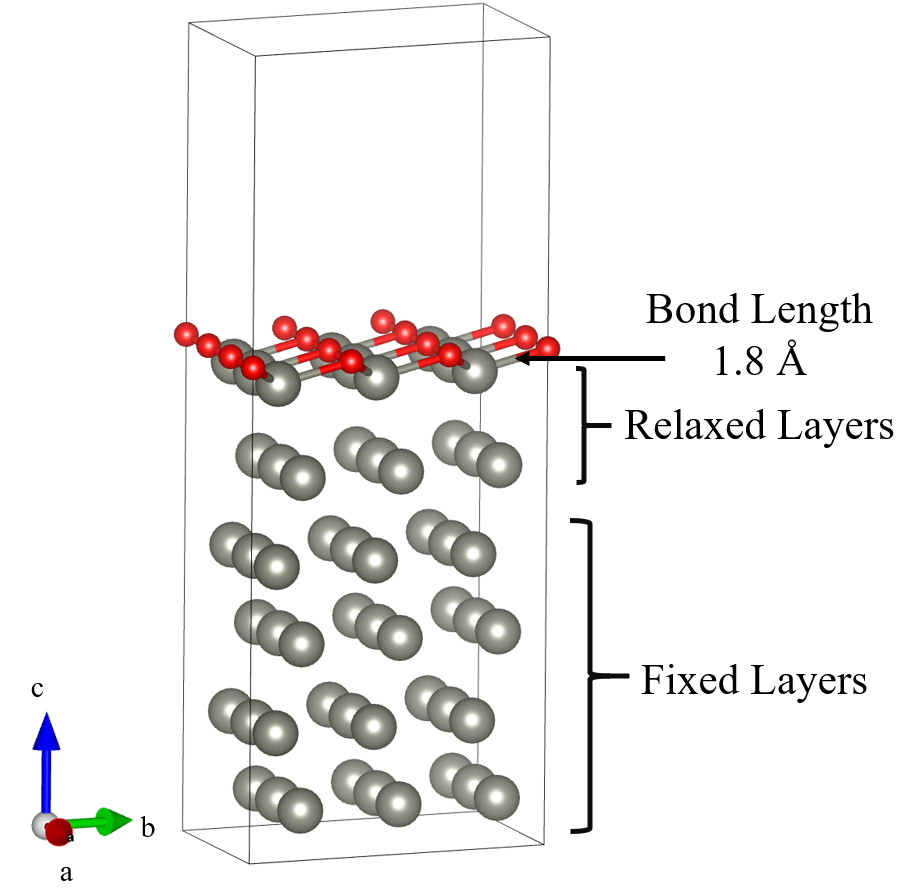}}
\qquad
\subfloat[]{
\label{}
\includegraphics[height=2.90in]{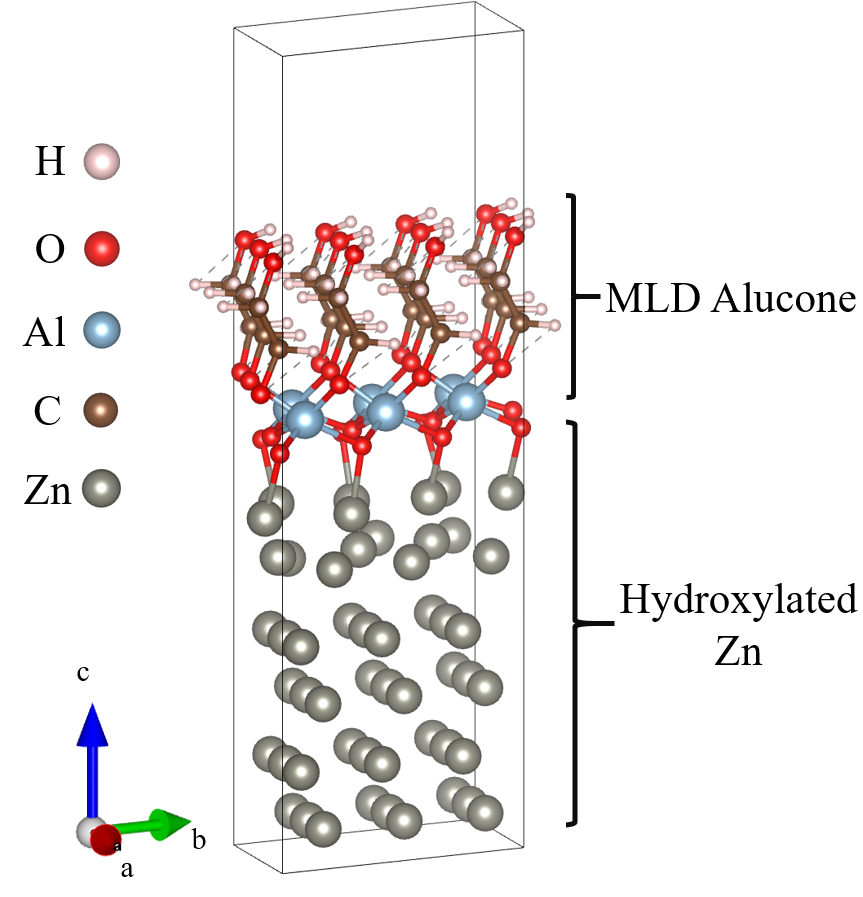}}
\qquad
\subfloat[]{
\label{}
\includegraphics[height=2.30in]{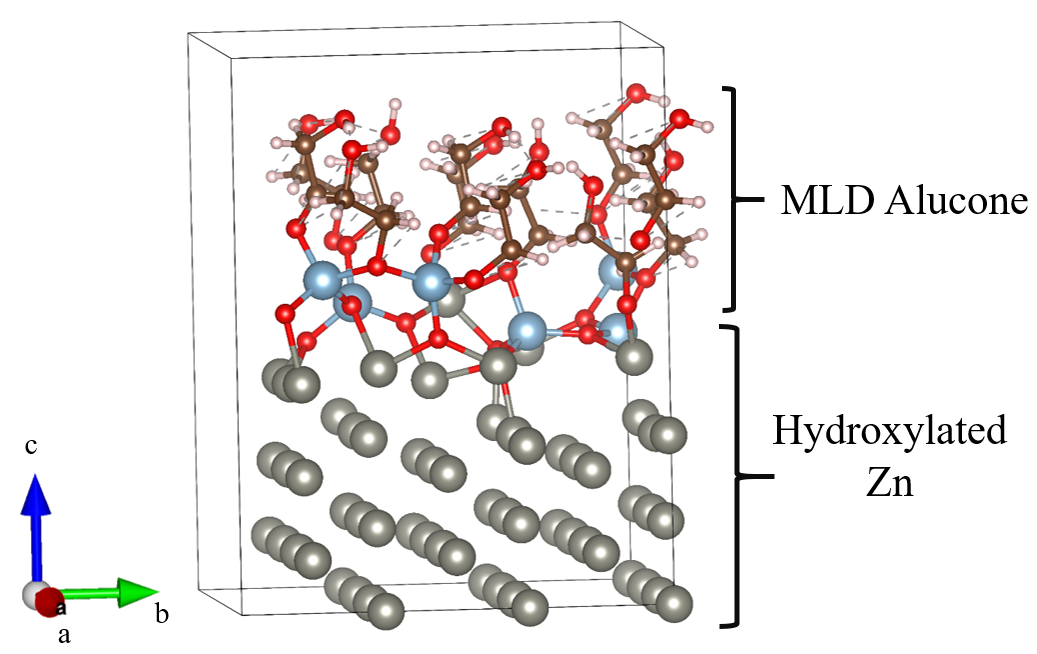}}
\caption {
(a) Optimized geometry of oxygen deposition in Zn \hkl(0001) surface using  binding sites of O atom in the middle of 6 Zn atoms (hollow-hexagonal close-packed (hcp)), (b) DFT optimized alucone thin film structure on hydroxylated Zn \hkl(0001) surface, and (c) (c) DFT optimized alucone thin film structure on hydroxylated Zn \hkl(1-100) surface. \label{fig:Schematic_alucone_MLD_growth}}
\end{figure}

\subsection{Chemical Stability Analysis}
From \tab{Adsorption_energy}, all three adsorption energies are negative for the adsorption reactions mentioned in equation \ref{adsorption_equations}. 
\begin{align}\label{adsorption_equations}
\mathrm{Zn_{slab}+ \left(\frac{n}{2}\right)O_{2}} &= \mathrm{Zn_{slab}-nO}\:\:(n=1)\nonumber \\
\mathrm{Zn_{slab}+ \left(\frac{n}{2}\right)H_{2} + \left(\frac{n}{2}\right)O_{2}} &= \mathrm{Zn_{slab}-nOH} \:\:(n=1)\nonumber\\
\mathrm{6Zn_{slab}-OH+ 6Al(CH_{3})_{3} + 9OHCH_{2}CH_{2}OH} &= \mathrm{6Zn_{slab}-O-AlOCH_{2}CH_{2}OH+15CH_{4}}
\end{align}
The negative adsorption energies for Zn-terminated surfaces indicate an exothermic process and the higher the adsorption energy indicates stronger interactions between adsorbates and the adsorbent. 
Therefore, the adsorption of alucone on hydroxylated Zn \hkl(0001) surface is thermodynamically stable. 
Furthermore, the range of the adsorption energies (-2.6 eV) indicate strong chemical adsorption of alucone group on the surface \cite{10.1016/j.physe.2012.11.020}.

\begin{table}[h]
\small
\setlength\tabcolsep{4pt}
 \begin{center}
\begin{tabular}{c c} 
\hline
  & eV per absorbate \\ [0.5ex]
 \hline
Oxygen deposited Zinc surface &  -0.8 \\ [1ex]
Hydroxylated Zinc surface &  -2.2  \\ [1ex] 
Alucone deposited hydroxylated Zinc surface (with vdW) & -2.6\\
 \hline
\end{tabular}
 \caption{\label{tab:Adsorption_energy} Adsorption energies of dffierent absorbate in Zn absorbant.}
\end{center}
\end{table}

\begin{figure}[h]
\centering
\subfloat[]{
\label{}
\includegraphics[width=0.40\linewidth]{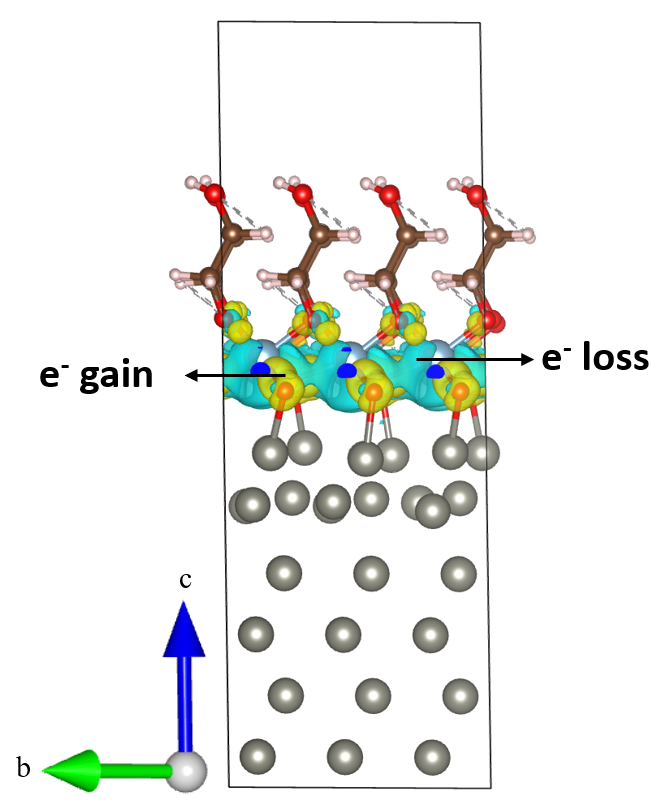}}
\qquad
\subfloat[]{
\label{}
\includegraphics[width=0.50\linewidth]{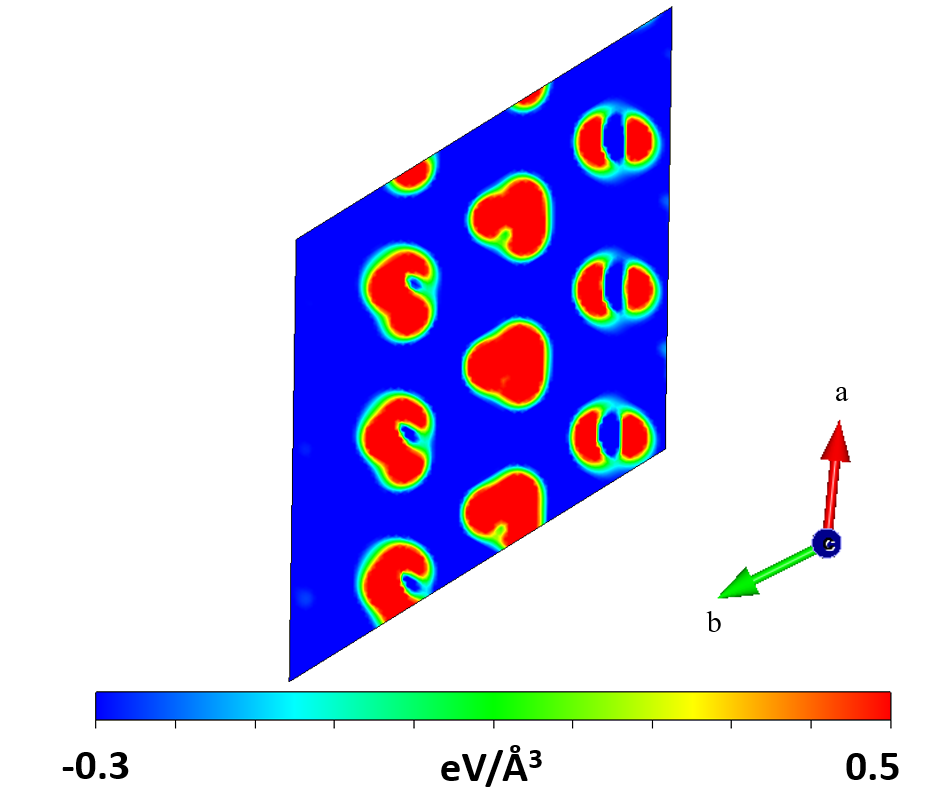}}
\qquad
\caption {
Differential charge density for the hydroxylated Zn with alucone.
Here, a, b and c are along \hkl[11-20], \hkl[2-1-10] and  \hkl[0001] direction, respectively.
(a) The yellow and blue regions in the isosurface show a gain and loss of electrons, respectively. 
(b) A slice along the ab plane where the red regions indicate the electron gains in oxygen on the Zn surface. 
The plotted charge density difference indicates the covalent nature that develops in between the absorbate and hydroxylated substrate.} \label{fig:charge_density_difference}
\end{figure}
\fig{charge_density_difference} maps the position of the redistributed electron densities for the MLD alucone adsorption on the hydroxylated Zn \hkl(0001) surface. 
Here, the loss of electrons is presented using blue zones whereas yellow zones point to electrons gain.
Due to the adsorption of alucone, the hydroxylated Zn surface displays a charge density increment (yellow isosurfaces) at the Zn-O bonds and a charge density reduction (blue isosurfaces) between the Al-O bonds (\fig{charge_density_difference} (a)). 
From the 2D slice along the ab plane (along with O atoms of the hydroxylated Zn surface) in  \fig{charge_density_difference}(b), the hydroxylated Zn surface  displays a sharp difference in charge density. 
Due to alucone deposition, there is a drastic rise in the charge density surrounding the hydroxylated surface O atoms (red regions). 
This sharp charge density increment indicates the covalent bonding behavior or chemisorption during the MLD deposition process. 
\begin{table}[h]
\small
\setlength\tabcolsep{4pt}
 \begin{center}
\begin{tabular}{l c c c c } 
\hline
Atom & Mean Bader Charge (e) & Electron Gain/Loss\\ \hline
$\textrm{Al}$ & $\text{2.4}$ & Loss\\
$\textrm{O} \text{ (hydroxylated Zn)}$ & $\text{-1.5}$ & Gain\\
$\textrm{O} \text{ (Alucone)}$ & $\text{-1.2}$ & Gain\\
$\textrm{Zn} \text{ (free layers)}$ & $\text{0.6}$ & Loss\\
$\textrm{C}$  & $\text{0.3}$ & Loss\\
$\textrm{H}$  & $\text{0.2}$ & Loss\\
 \hline
\end{tabular}
 \caption{\label{tab:bader_charge}Bader net atomic charges.}
\end{center}
\end{table}

Unlike charge-density difference, which is a qualitative analysis technique, Bader charge analysis can present the extent of the chemical reaction between atoms (hydroxyl groups on precursor decomposition) quantitatively \cite{10.1088/0953-8984/21/8/084204}.
\tab{bader_charge} summarizes the net Bader atomic charges for each atom. 
From~\tab{bader_charge}, it is logical that chemisorption dominates the adsorption process due to charge transfer between absorbate and absorbent. 
The sum of the Bader charges of O and H atom has to equal that of the central Al atom for complete physisorption.
In that case, only weak van der Waals forces could be applicable between the precursor and the surface. 
However, for the alucone coated hydroxylated Zn \hkl(0001) surface, we observed the opposite trend here. 
The Al atom has more positive (loss of electron) net atomic Bader charges after adsorption while the charges increase for O atoms. 
Therefore, the Al-hydroxylated Zn surface bonds get stronger which indicates chemisorption due to the changes in the surface electronic properties. 
\subsection{Mechanical Stability  and Properties}
\begin{figure}[h]
\centering
\subfloat[]{
\label{}
\includegraphics[height=2.2in]{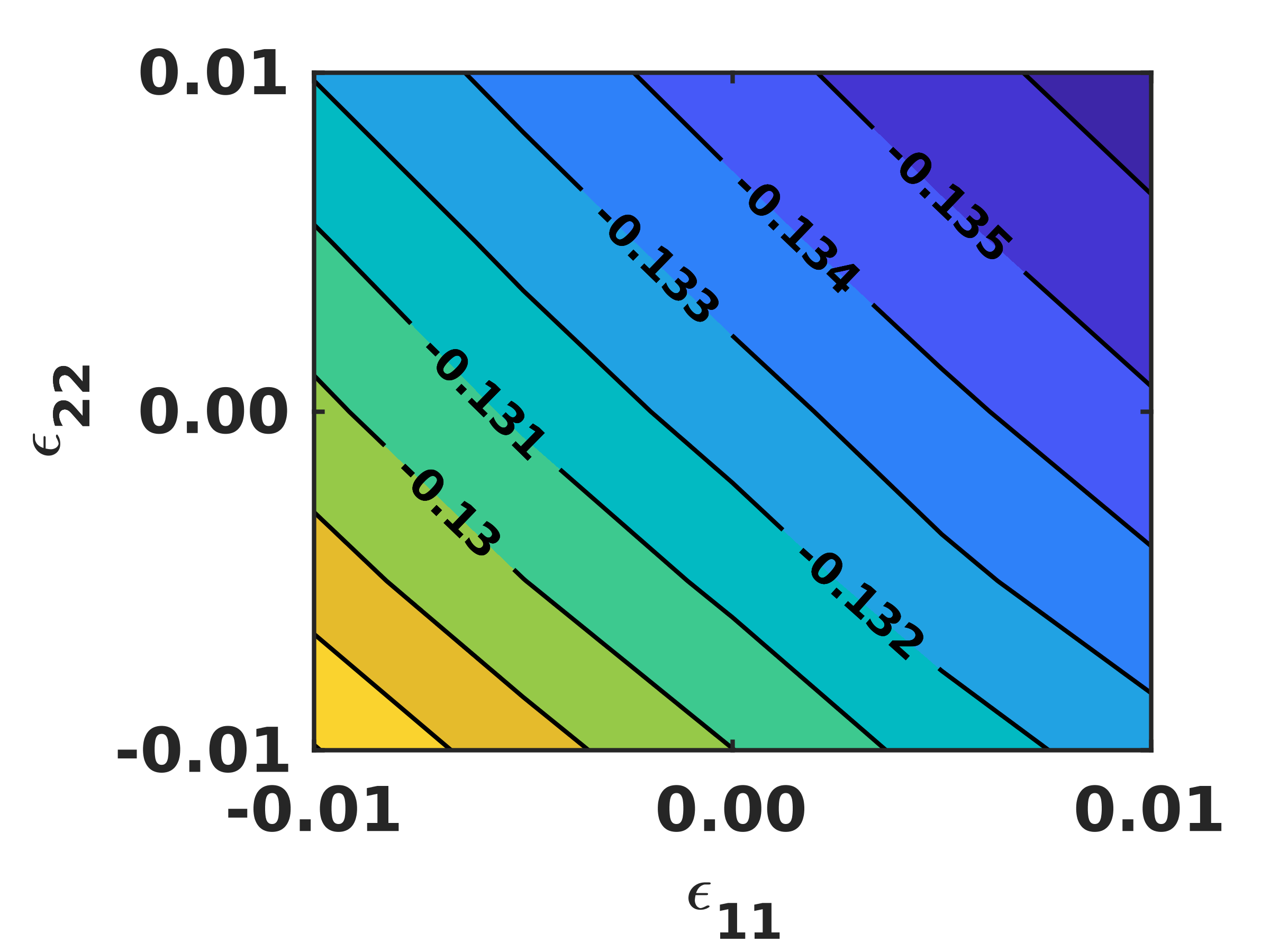}}
\qquad
\subfloat[]{
\label{}
\includegraphics[height=2.2in]{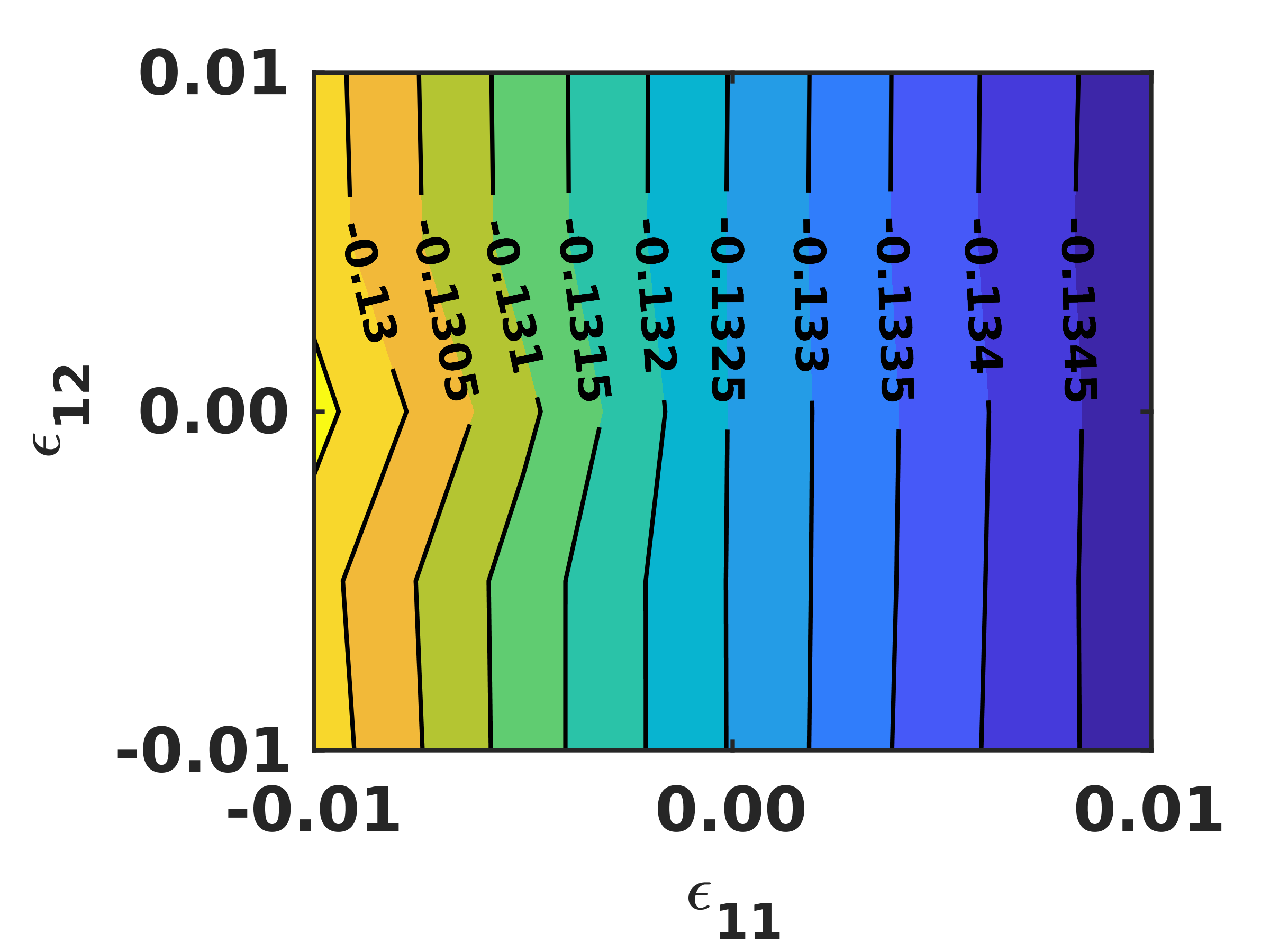}}
\qquad
\subfloat[]{
\label{}
\includegraphics[height=2.2in]{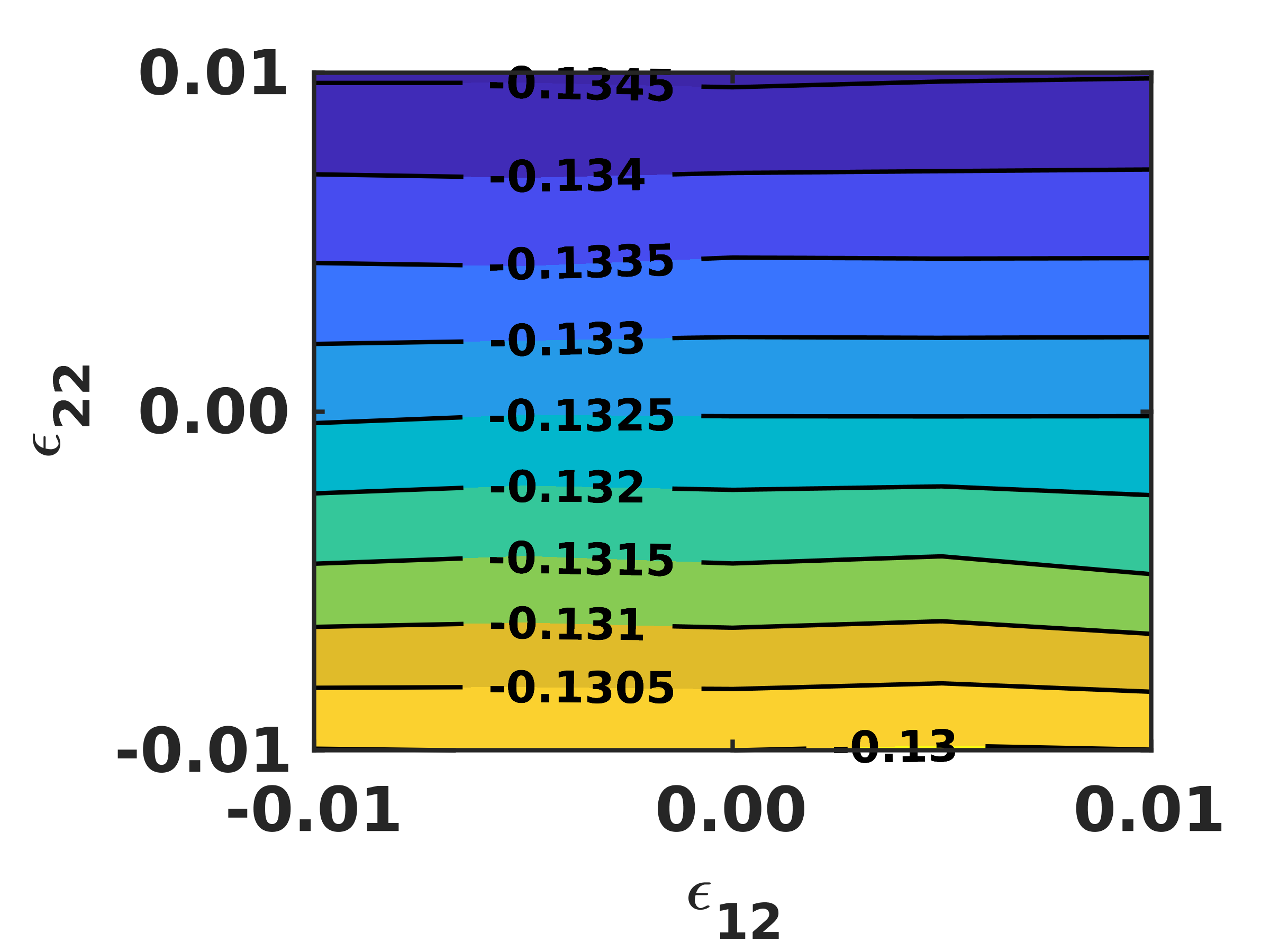}}
\qquad
\caption {\label{fig:surface_energy_with_strain}Contour plot of variation of surface energy ($\mathrm{eV/\textrm{\r{A}}^{2}}$) with biaxial strain for alucone coated hydroxylated Zn \hkl(0001) surface (a) 11-22, (b) 11-12, and (c) 12-22 direction.}
\end{figure}
The surface energy of alucone-coated Zn (\fig{surface_energy_with_strain} and \fig{surface_energy_with_strain_prismatic_alucone}) and hydroxylated Zn surface is negative with the strain which is possible for multi-component systems.
But surface energy in a single-component stable solid such as Zinc is positive (\fig{surface_energy_with_strain_bulk_Zn_ZNOH}) \cite{10.1038/nmat1336a, 10.1038/nmat1106}. 
That is because a clean solid surface's interfacial energy stays in equilibrium with its own vapor and therefore no chemical effects for changing the Gibbs free energy.
The change in Gibbs free energy of the substrate per unit surface area at constant temperature and pressure defines the surface energy of the system. 
However, for multi-component system (such as Alcuone on hydroxylated Zn surface) chemical effects must be taken into consideration which changes the solid's surface energy. 
Furthermore, adsorption is an exothermic process on solid surfaces which reduces the surface energy of the solid.
Large adsorption energies as shown in~\tab{Adsorption_energy} also lead to negative surface energies.
From~\fig{surface_energy_with_strain}, surface energy raises with tensile strain but decreases by compressive strain.
This phenomenon leads to the roughness of the strained layer which altered the surface energy~\cite{10.1103/PhysRevLett.74.4962}. 
Therefore, roughness has been promoted by compressive strain whereas it has been inhibited by tensile strain which is also observed by Xie \emph{et al.} \cite{10.1103/PhysRevLett.73.3006, 10.1103/PhysRevLett.74.4962}. 
\begin{table}[h]
\small
\setlength\tabcolsep{4pt}
 \begin{center}
\begin{tabular}{c c c c c } 
\hline
 & $\gamma^{0}$ & $\tau^{0}_{11}$  & $\tau^{0}_{22}$ & $\tau^{0}_{12}$ \\ [0.5ex]
 \hline
Zinc surface & 42 & -96 & -95 & 0 \\ [1ex]
Hydroxylated zinc surface &  -151 & -283 & -321 & 0\\ [1ex] 
Alucone deposited zinc \hkl(0001) surface & -133 & -373 & -374 & -3 \\ [1ex]
Alucone deposited zinc \hkl(1-100) surface & -225 & -400 & -398 & 0\\
 \hline
\end{tabular}
 \caption{\label{tab:surface_stress} List  of surface stress $(\tau)$ ($\mathrm{meV/\textrm{\r{A}}^{2}}$) and surface energy $(\gamma)$  ($\mathrm{meV/\textrm{\r{A}}^{2}}$) of zinc, oxygen deposited zinc and alucode deposited zinc thin films.} 
\end{center}
\end{table}
From our optimized alucone thin film, there is a lattice mismatch with the substrate which rises compressive strain that induces surface reconstruction.
The Zn metal surface observes an increased atom number than bulk due to surface reconstruction and hence exhibits negative surface stresses as shown in~\tab{surface_stress}.
However, negative surface stress makes a surface unstable\cite{10.1080/01418618608242900}, unless its absolute magnitude of it is small~\cite{10.1016/0039-6028(91)90269-X}.
Impacts of surface instabilities due to large surface stress can be manifold. 
Firstly, compressive (negative) surface stress can result in buckling of the surface which is well known for the unstable elastic continuum model~\cite{10.1063/1.324298}.
Secondly, atoms' density within the surface layer may change due to this instability. 
From an atomistic viewpoint, there are three effects related to the variation in the density of surface atoms due to the surface reconstructions \cite{10.1088/0953-8984/1/41/006,10.1016/0039-6028(91)90269-X}. 
Due to inter-atomic interactions, the natural bond lengths change compared to bulk systems. 
During reconstruction, the extra energy cost is associated with the atom transfer from or to the surface layers.
Finally, the surface and substrate atoms' interaction changes due to possible disturbance of the surface-substrate bonding.
\begin{table}[h]
\small
\setlength\tabcolsep{4pt}
 \begin{center}
\begin{tabular}{c c c c c c c c c c} 
\hline
  & $C^{{\textrm{S}}}_{1111}$  & $C^{{\textrm{S}}}_{1122}$ & $C^{{\textrm{S}}}_{1212}$ & $C^{{\textrm{S}}}_{2222}$ & $C^{{\textrm{S}}}_{2211}$ & $C^{{\textrm{S}}}_{1112}$ & $C^{{\textrm{S}}}_{1222}$ & $C^{{\textrm{S}}}_{2212}$ &  $C^{{\textrm{S}}}_{1211}$\\ [0.5ex]
 \hline
Zinc  & -1.5 & 11.4 & -6.4 & 0.0 & 0.0 & 0.0 & 0.0 & 0.0 & 0.0 \\ [1ex]
Hydroxylated Zinc  & 8.7 & 0.4 & 0.0 & 6.8 & 0.4 & 0.0 & 0.0 & 0.0 & 0.0\\ [1ex] 
Alucone deposited Zinc \hkl(0001) & 2.1 & 2.1 & 0.3 & 8.5 & 2.1 & -1.2 & 1.4 & 1.4 & -1.2 \\ [1ex] 
Alucone deposited Zinc \hkl(1-100) & 5.8 & 1.7 & 0.0 & 6.8 & 1.7 & 0.0 & 0.0 & 0.0 & 0.0 \\
 \hline
\end{tabular}
 \caption{\label{tab:surface_elatic_constant} List  fourth-order surface elastic-stiffness coefficients $(C^{\textrm{S}}_{ijkl})$ ($\mathrm{eV/\textrm{\r{A}}^{2}}$) of zinc, hydroxylated deposited zinc and alucode deposited zinc surface.}
\end{center}
\end{table}

\par Point to be noted that $C^{\textrm{S}}_{ijkl}$ not necessarily is always positive like the bulk Zn elastic tensor (\tab{zinc_crystal_elastic_constants}).
It seemingly contradicts the postulates of basic thermodynamics which assures the stability of the solid on the bulk elastic modulus tensors' positive definiteness.
However, it is not applicable for the surface elastic tensor $C^{\textrm{S}}_{ijkl}$. 
The reason behind this is that a surface cannot remain individually without the bulk and therefore, only the combined bulk-surface model requires to fulfill the positive definiteness criterion~\cite{10.1103/PhysRevB.71.094104}.
It should be pointed out that the surface elastic constants ($C^{\textrm{S}}_{ijkl}$) have dimensions of force$\cdot$length$^{-1}$ (e.g., N/m or eV/\r{A}$^{2}$), which is dissimilar to the bulk elastic constants, as $C^{\textrm{S}}_{ijkl}$ exists on a two-dimensional surface.
It is well accepted that the easiest slip for HCP materials such as zinc is along with the basal slip system \hkl[11-20]\hkl(0001), whereas slip is notably harder along other plane directions (such as prismatic and pyramidal planes) due to higher-order stress requirements~\cite{10.1016/j.jmps.2017.12.009}.
In our study, we modeled prismatic \hkl(1-100) surfaces to understand the impact of plane orientation on surface properties. From \tab{surface_elatic_constant}, it is clear that crystalline orientation plays a significant role to determine the surface elastic constants, which contributes to the anisotropic elastic deformation of the system.
\subsubsection{Mechanical Figures of Merits of Alucone-Coated Thin Films} 
\begin{table}[h]
\small
\setlength\tabcolsep{4pt}
 \begin{center}
\begin{tabular}{c c c c c c c c c c } 
\hline
 & $C_{11}^{\textrm{film}}$  & $C_{22}^{\textrm{film}}$ & $C_{12}^{\textrm{film}}$ & $C_{21}^{\textrm{film}}$ & $C_{13}^{\textrm{film}}$ & $C_{23}^{\textrm{film}}$ & $C_{31}^{\textrm{film}}$ & $C_{32}^{\textrm{film}}$ & $C_{33}^{\textrm{film}}$\\ [0.5ex]
\hline
Zinc  & 141.3 & 149.5 & 1.4 & 0.0 & 0.0 & 0.0 & 0.0 & 0.0 & 0.0\\[1ex] 
Hydroxylated Zinc & 149.2 & 147.6 & 5.9 & 5.9 & 0.0 & 0.0 & 0.0 & 0.0 & 67.5 \\ [1ex] 
Alucone coated Zinc \hkl(0001) & 143.6 & 147.7 & 6.9 & 6.9 & 0.0 & 0.0 & 0.0 & 0.0 & 67.5 \\ [1ex] 
Alucone coated Zinc \hkl(1-100) & 142.6 & 142.7 & 5.7 & 5.7 & 0.0 & 0.0 & 0.0 & 0.0 & 67.5  \\
\hline
\end{tabular}
 \caption{\label{tab:plane_stress_elastic_constant} List  of second-order effective elastic-stiffness coefficients $(C_{ij}^{\textrm{film}})$ (GPa) of different thin films.}
\end{center}
\end{table}

\par A film thickness of $L_{z}=50$~nm (see~\eq{eq11}) was used to calculate the mechanical properties of different thin films, which satisfies the positive definiteness condition of all elastic modulus tensors (\fig{size_dependent_elastic_properties}). 
From~\tab{plane_stress_elastic_constant}, $C_{ij}^{\textrm{film}}$ shows the thin film's stiffness level to the bulk material. 
For bulk Zn, planar elastic constants are less than the bulk material which indicates that the bare Zn thin film is softer than its bulk counterpart. 
\begin{table}[h]
\small
\setlength\tabcolsep{4pt}
 \begin{center}
\begin{tabular}{ c c c c c c} 
\hline
& B & G & E & $\nu$ & $k$ \\ [0.5ex]
\hline
Zinc  & 16.7 & 13.4 & 31.6 & 0.18 & 1.2\\ [1ex] 
Hydroxylated Zinc & 18.6 & 14.5 & 34.5 & 0.19 & 1.3 \\ [1ex] 
Alucone coated Zinc \hkl(0001) & 18.6 & 11.4 & 28.5 & 0.24 & 1.6 \\ [1ex] 
Alucone coated Zinc \hkl(1-100) & 17.8 & 14.4 & 34.0 & 0.18 & 1.2 \\
\hline
\end{tabular}
 \caption{\label{tab:plane_stress_modulus} List  of Young's modulus (E) (GPa), Shear modulus (G) (GPa), Bulk modulus (B) (GPa), Poisson's ratio \((\nu)\) and Pugh ratio (k) of different thin films.}
\end{center}
\end{table}

The Poisson's and Pugh's ratios can be evaluated as the benchmark of ductility. 
Dimensionless quantities like Poisson's and Pugh's ratio suggest a direct method of anticipating ductility from first principles as they are determined from bulk (B) and shear modulus (G).
From~\tab{plane_stress_modulus}, alucone-coated hydroxylated basal Zn thin film has a Pugh's ratio close to 1.75~\cite{Michael} and Poisson's ratio close to 0.26~\cite{Michael}, suggesting a good ductility of the thin film. 
However, for the prismatic surface \hkl(1-100), the Pugh and Poisson's ratio are less than the cutoff values for ductility, due to the increase of G and decrease of B compared to the \hkl(0001) surface. 
This observation suggests that the coating could lead to an increased anisotropic deformation mode in Zn, facilitating twinning as in other hcp metals\cite{10.1016/j.jmps.2017.12.009}.
These thin film properties are size-dependent as the contribution of surface elastic constants on thin film elastic constants decreases exponentially with increasing film thickness (\fig{size_dependent_elastic_properties}).
Alucone-coated Zn met all possible necessary and
sufficient elastic stability criteria: $\textrm{det(C}_{ij})>0$; $\lambda\mathrm{_{1}}>0,\lambda\mathrm{_{2}}>0,\lambda\mathrm{_{3}}>0$; leading principle minors $>0$ and trailing minors $>0$~\cite{Mouhat, 10.1088/2053-1583/ab2ef3}. 
Here, $\lambda$ indicates the eigenvalues of second order elastic constant matrix $\mathrm{C}$.
\par From~\tab{toughness_hardness_membrain_strain}, the estimated hardness ($H_\mathrm{V}$) of the alucone-coated thin film in the basal surface \hkl(0001) is less than the uncoated thin film. 
However, the hardness of the prismatic  surface \hkl(1-100) is higher than the basal surface.
From~\tab{toughness_hardness_membrain_strain}, fracture toughness of alucone deposited Zn \hkl(0001) thin film was estimated to be $\sim0.5$ MPa$\cdot$m$\mathrm{^{1/2}}$. 
From previous experimental measurements with a fracture mechanics model using crack density versus applied tensile strain, the $\mathrm{{\it{K}}_{IC}}$ of alucone films on polyethylene naphthalate (PEN) substrate was determined to be 0.16-0.18 MPa$\cdot$m$\mathrm{^{1/2}}$\cite{10.1063/1.3124642}, which is in the same order of magnitude as our prediction.
In the case of surface wrinkling, the compressive stress in thin film deposition is generated by mechanical force and temperature mismatch. 
The membrane strain has to be small for surface wrinkling and for 1D wrinkling in alcuone deposited Zn thin film, we observed a membrane strain of 0.5-0.6 \%. 
\begin{table}[h]
\small
\setlength\tabcolsep{4pt}
 \begin{center}
\begin{tabular}{c c c c c} 
\hline
 & $H_\mathrm{V}$  & $\mathrm{K_{\textrm{IC}}}$  & $\epsilon_{\textrm{m}}$ & $\sigma_{\textrm{m}}$\\ [0.5ex]
 \hline
Zinc thin film & 3.3 & 0.3 & 0.0 & 0.0 \\ [1ex]
Hydroxylated zinc thin film & 3.4 & 0.4 & -0.5 & -16.9\\ [1ex] 
Alucone deposited zinc \hkl(0001) thin film & 2.1 & 0.5 & -0.6 & -15.6\\ [1ex]
Alucone deposited zinc \hkl(1-100) thin film & 3.5 & 0.6 & -0.5 & -16.8\\ 
 \hline
\end{tabular}
\caption{\label{tab:toughness_hardness_membrain_strain} List  of Vickers hardness, $H_\mathrm{V}$ (GPa), fracture toughness, $\mathrm{K_{\textrm{IC}}}$ (MPa$\cdot$m$\mathrm{^{1/2}}$) and threshold membrane strain ($\epsilon_{\textrm{m}}$ in percentage) of zinc, hydroxylated zinc, and alucode deposited zinc thin films.}
\end{center}
\end{table}
The development of in-plane compressive stress is one of the main causes of dendrite formation in the thin film deposition technique~\cite{10.1007/s10483-020-2596-5}.
Two cases are possible due to this compressive stress. First case: when compressive deposition stress develops in alucone film, it will be transported to the Zn substrate.
As a consequence, it causes Zn to wrinkle above a threshold membrane strain. 
Second case: the developed stress cannot be relaxed and eventually dendrites grow in the Zn substrate. 
The compressive stress in MLD alucone thin film due to wrinkling is evaluated to be $\sigma_{\textrm{m}}=E_{\textrm{alucone}}*\epsilon_{\textrm{m}}$~\cite{10.1038/s41560-018-0104-5} by employing our calculated $\mathrm{{\it{E}}_{alucone}}$. 
This stress is well below the microsized alucones' yield strength and the compressive residual stress level~\cite{10.1016/j.actamat.2009.07.015}. 
The above estimation suggests that the alucone-coated Zn anode is  dendrite free and likely to have wrinkles.
\section{Conclusion}

\par Using a combination of chemo-mechanical analysis, we explored the synergy between alucone coating's stability onto Zn. 
We analyzed the mechanistic insights into the alucone's chemo-mechanical stability due to the molecular layer deposition technique. 
We studied the alucone precursor and hydroxylated Zn surface interactions by quantifying their charge density difference, Bader charges, and adsorption energies. 
Coupling the high adsorption energies with the charge transfer analysis, namely Bader charge analysis (quantitative) and charge density difference (qualitative), demonstrated that chemisorption is the dominant mechanism of MLD alucone deposition.
We found out the presence of compressive stress and wrinkling formation during coating deposition and its role in the stability of alucone coating. 
A critical analysis of the surface stress and surface energy level predicted surface reconstruction which induced residual stress during alucone coating on the Zn substrate. 
The common fact for the thin film in different Zn planes \hkl(0001)/\hkl(1-100) was that all the coated surfaces have negative surface elastic constants and Young modulus was less than that of the bulk counterpart. 
Moreover, our study offers a systematic framework to analyze the chemo-mechanical stability of coatings composed of a new class of material for the development of dendrite-free electrodes for other metallic electrode systems.
The distribution and effects of the residual stresses imposed on the Zn anodes by the coating deposition need to be extensively studied in the future to develop strategies to utilize dendrite mitigation techniques for feasible AZIBs.

\section{Acknowledgement}
We acknowledge the support of New Frontiers in Research Fund (NFRFE-2019-01095) and from the Natural Sciences and Engineering Research Council of Canada (NSERC) through the Discovery Grant under Award Application Number 2016-06114. M.G. gratefully acknowledges the financial support from the Department of Mechanical Engineering at UBC through the Four Years Fellowship. 
This research was supported through high-performance computational resources and services provided by Advanced Research Computing at the University of British Columbia and Digital Research Alliance of Canada.

\providecommand{\latin}[1]{#1}
\makeatletter
\providecommand{\doi}
  {\begingroup\let\do\@makeother\dospecials
  \catcode`\{=1 \catcode`\}=2 \doi@aux}
\providecommand{\doi@aux}[1]{\endgroup\texttt{#1}}
\makeatother
\providecommand*\mcitethebibliography{\thebibliography}
\csname @ifundefined\endcsname{endmcitethebibliography}
  {\let\endmcitethebibliography\endthebibliography}{}

\clearpage
\onecolumngrid
\raggedbottom
\setcounter{section}{0}
\setcounter{equation}{0}
\setcounter{figure}{0}
\setcounter{table}{0}
\setcounter{page}{1}
\makeatletter
\renewcommand{\thesection}{S\arabic{section}}
\renewcommand{\theequation}{S\arabic{equation}}
\renewcommand{\thefigure}{S\arabic{figure}}
\renewcommand{\thetable}{S\arabic{table}}
\renewcommand{\thepage}{SM\arabic{page}}

\renewcommand{\bibnumfmt}[1]{[S#1]}
\renewcommand{\citenumfont}[1]{S#1}

\begin{center}
{\Large \bf Supporting Information}
\end{center}

\section{Computational Details}\label{sec:SI-CompDet}

\subsection{Determining Fixed and Free Layers of Zn Thin Film Model}
\begin{figure}[h]
\centering
\subfloat[]{
\label{}
\includegraphics[height=2.18in]{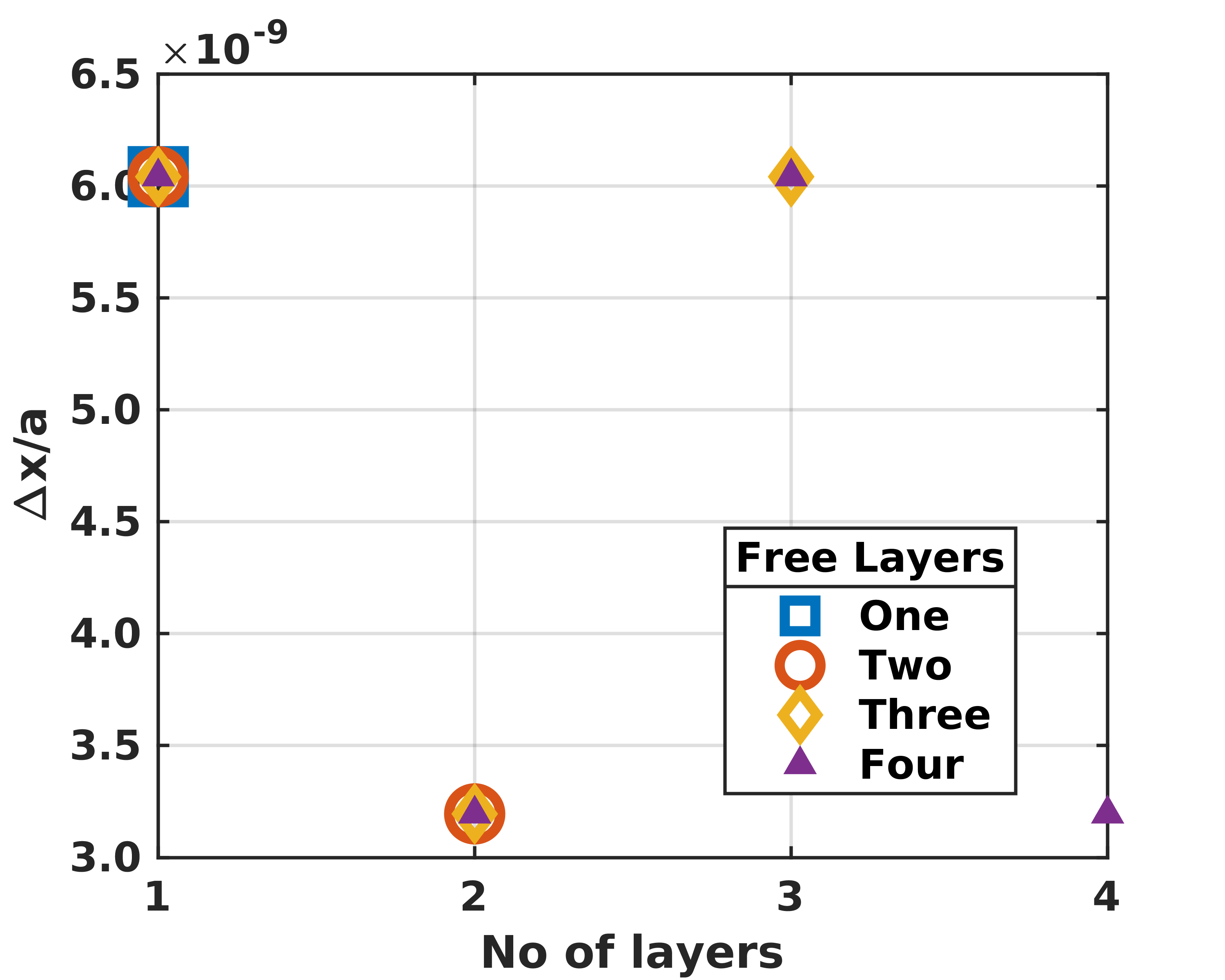}}
\qquad
\subfloat[]{
\label{}
\includegraphics[height=2.18in]{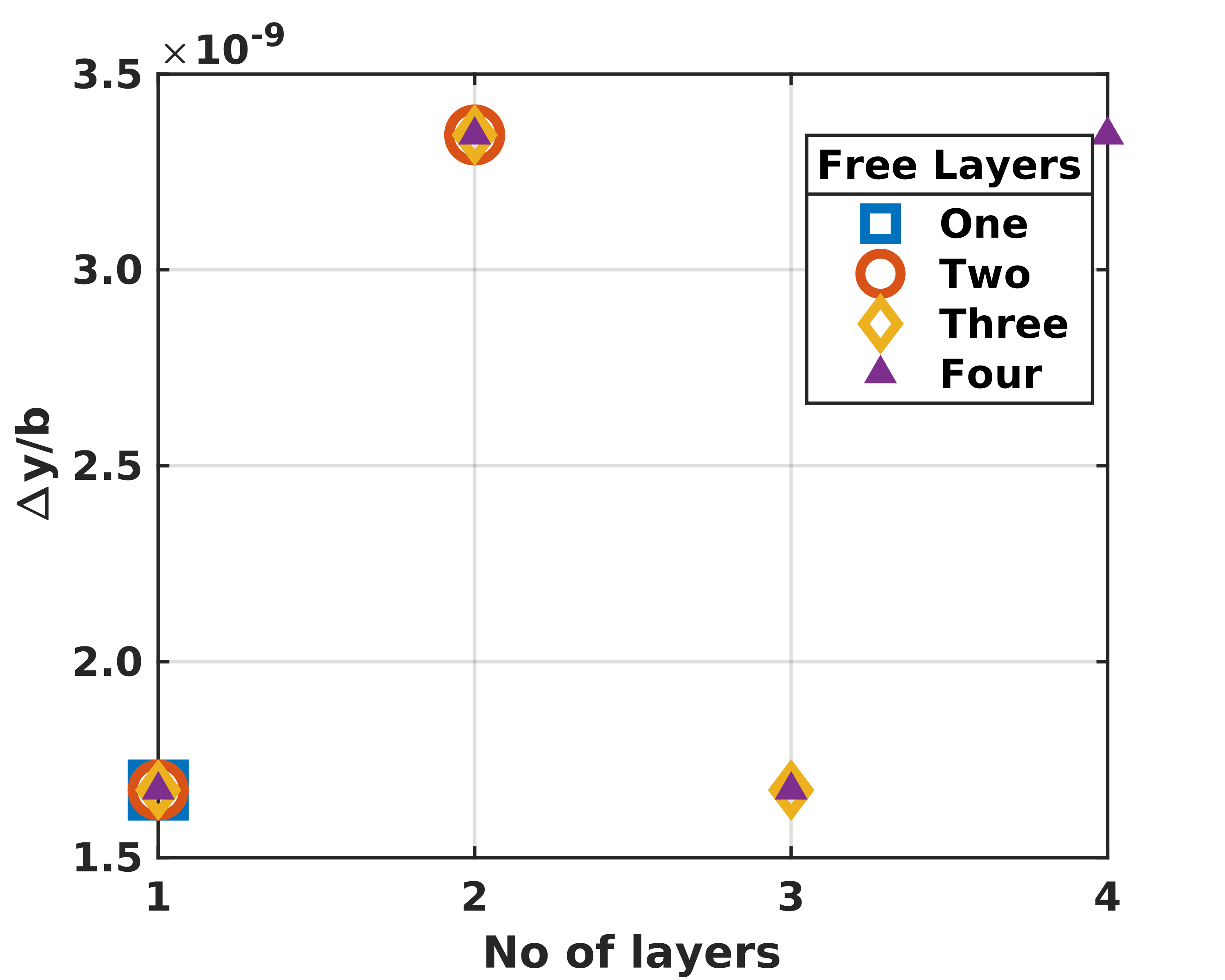}}
\qquad
\subfloat[]{
\label{}
\includegraphics[height=2.18in]{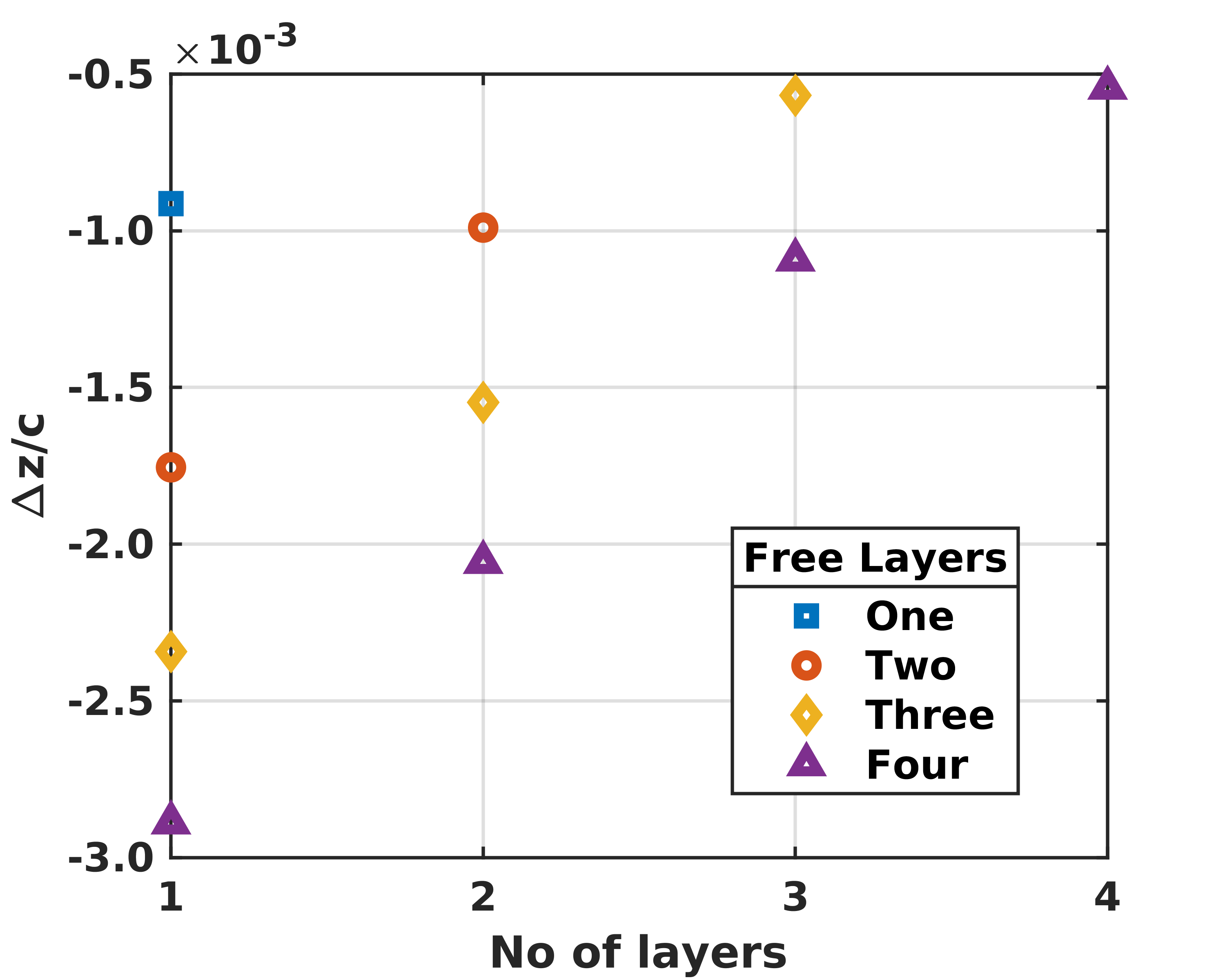}}
\qquad
\subfloat[]{
\label{surface_relaxation_no_of_free_layer}
\includegraphics[height=2.18in]{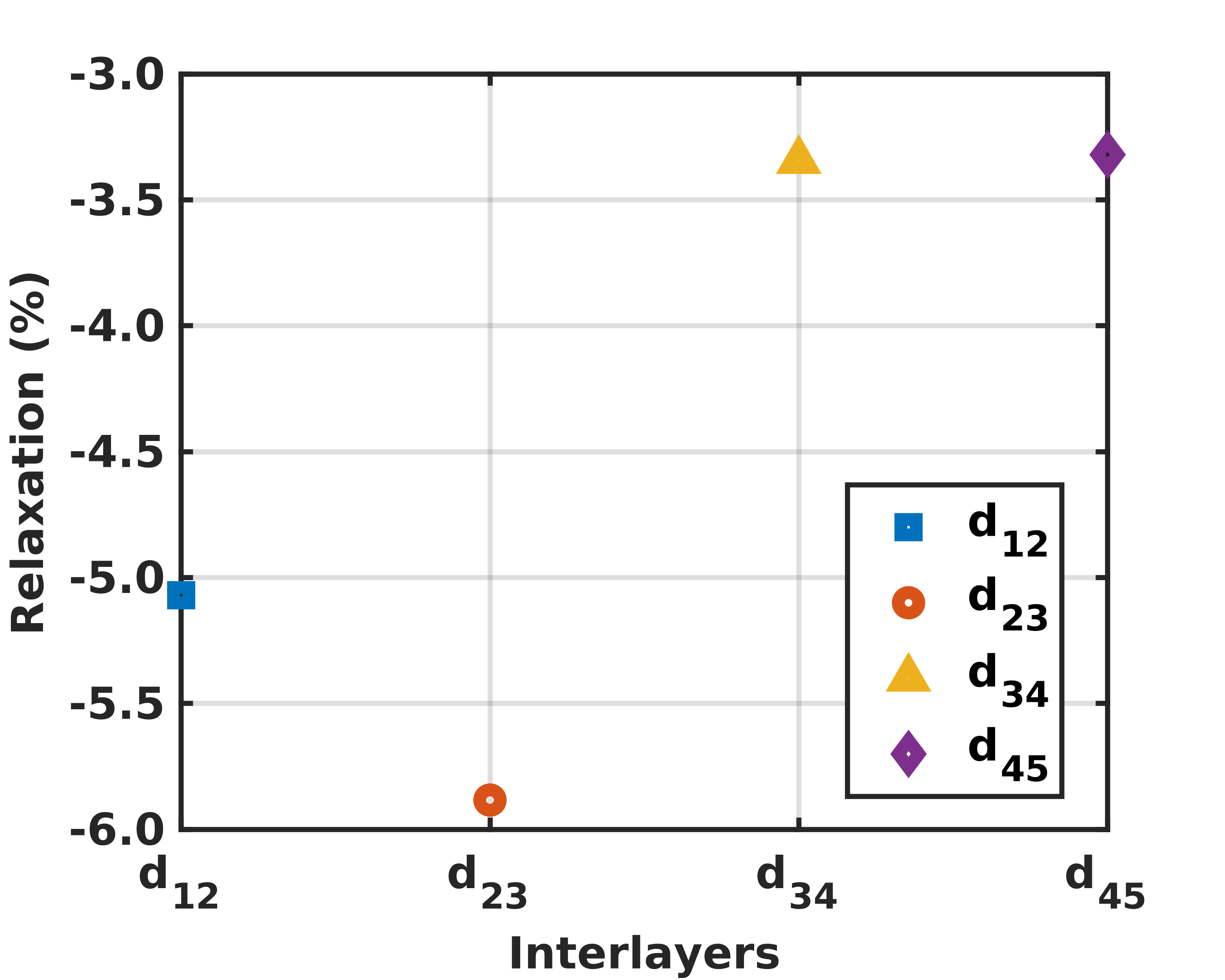}}
\caption {
The relative movement of free surface layers with respect to lattice parameters in fractional co-ordinates in (a) x, (b) y, and (c) z-direction, and (d) Relaxation estimation, \(d_{ij}\) (i,j=1,2,3,$\ldots$) as a percentage of the interlayer distance of bulk for the Zn \hkl(0001) thin film.
Here, positive and negative values specifies an expansion and contraction, respectively. 
We varied the number of free layers to determine optimal number of free layers.\label{fig:no_of_free_layer_optimization}}
\end{figure}

From~\fig{no_of_free_layer_optimization} (a, b, c), the computational analysis indicated that relaxations primarily existed in the z-direction.  
Surface atoms do not reconstruct in x and y-direction as negligible in-plane movements were noticed. 
The level of relaxation was found by computing the distance between the \(i\: \&\: j\) layer shifts which we presented here as interlayer distance (\(d_{ij}\)) .
The upwards shifting  of atoms towards the surface results in a positive value, whereas a downwards shifting in the direction of the bulk results in a negative value of \(d_{ij}\).
We found that the hexagonal close-packed (hcp) Zn metal exhibits large inward layer relaxation shown in~\fig{no_of_free_layer_optimization} (c,d). 
\begin{table}[H]
\begin{center}
\begin{tabular}{c c c c c c} 
\hline
& & & layer & relaxation &\\ [0.5ex]
\cline{3-6}
& Surface & \(d_{12}\) & \(d_{23}\) & \(d_{34}\) & \(d_{45}\)\\ [0.5ex]
\hline
Zn & 0 0 0 1 & -5.1 & -5.9 & -3.3 & -3.3\\  [1ex] 
\hline
\end{tabular}
\caption{\label{tab:no_of_free_layer_optimization}  Calculated surface layer relaxations (\(d_{ij}\) in percent) as a percentage of the interlayer distance of bulk for the top four layers of Zn \hkl(0001) thin film.}
\end{center}
\end{table}
In \tab{no_of_free_layer_optimization}, the layer relaxation is interpreted as ~\cite{Zolyomi_V} - 
\begin{equation}
d_{ij} = (\lambda_{{ij}}^{\textrm{s}} - \lambda_{ij}^{\textrm{b}}) / \lambda_{ij}^{\textrm{b}}
\end{equation} 
where \(\lambda_{ij}^{\textrm{b}}\) and \(\lambda_{ij}^{\textrm{s}}\) are the bulk and surface interlayer spacing, respectively.
The relaxations in multilayer Zn metal indicate that  the relaxation reduces in the z-direction with the depth from the surface. 
\par Our calculations indicate the movement of the top four layers with an 8-layer Zn thin film model. 
The distance the first four layers shifted was -0.4, -0.5, -0.3, and -0.3 pm, respectively. 
The contraction noted in our calculations for all four free layers of the zinc surfaces, can be interpreted using the electrostatic model~\cite{Spencer_Michelle_J_S}. 
This model expresses that the outermost surface atoms followed an inward movement for reorganizing themselves to acquire a consistent electron density parallel to the Zn surface. 
As a consequence, atoms in the lower layers rearranged their positions to adjust neighboring displacements. 
The contraction phenomenon of the free atomic layers can also be explained using the idea of the bond-order/bond-length.
The surface atomic bonds are broken during relaxation and electrons relocate to form a stronger bond which results in a small bond length in the top two free layers of  the Zn surface model~\cite{Spencer_Michelle_J_S}. 
This can be validated by the relaxations of the lower layers of our model where relaxation decreases and swiftly advance the bulk spacings (-5.1\% to -3.3\%), indicating the model is acceptable for the study of zinc surface with two free surface layers.
\begin{figure}[H]
\centering
\subfloat[]{
\label{}
\includegraphics[height=2.18in]{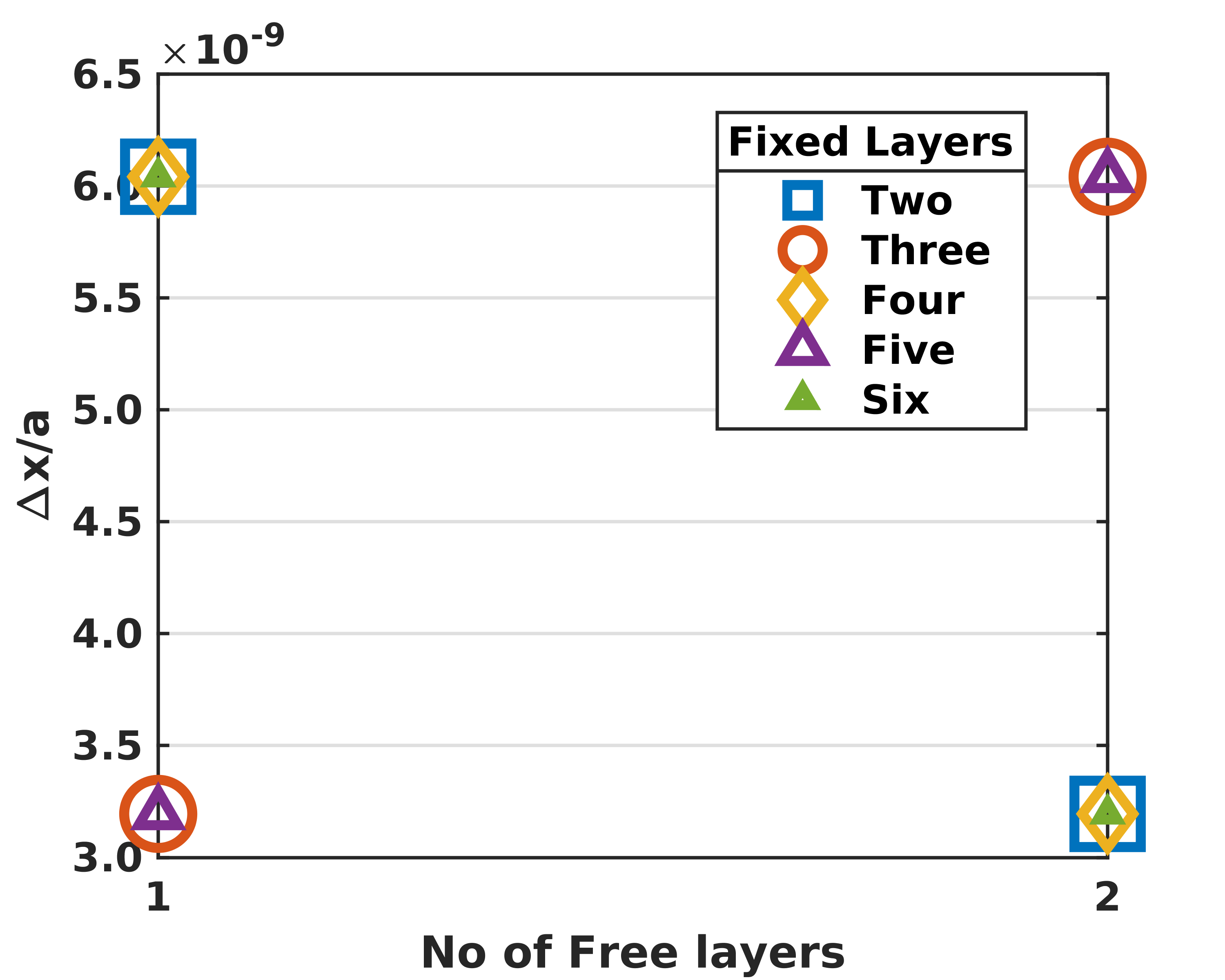}}
\qquad
\subfloat[]{
\label{}
\includegraphics[height=2.18in]{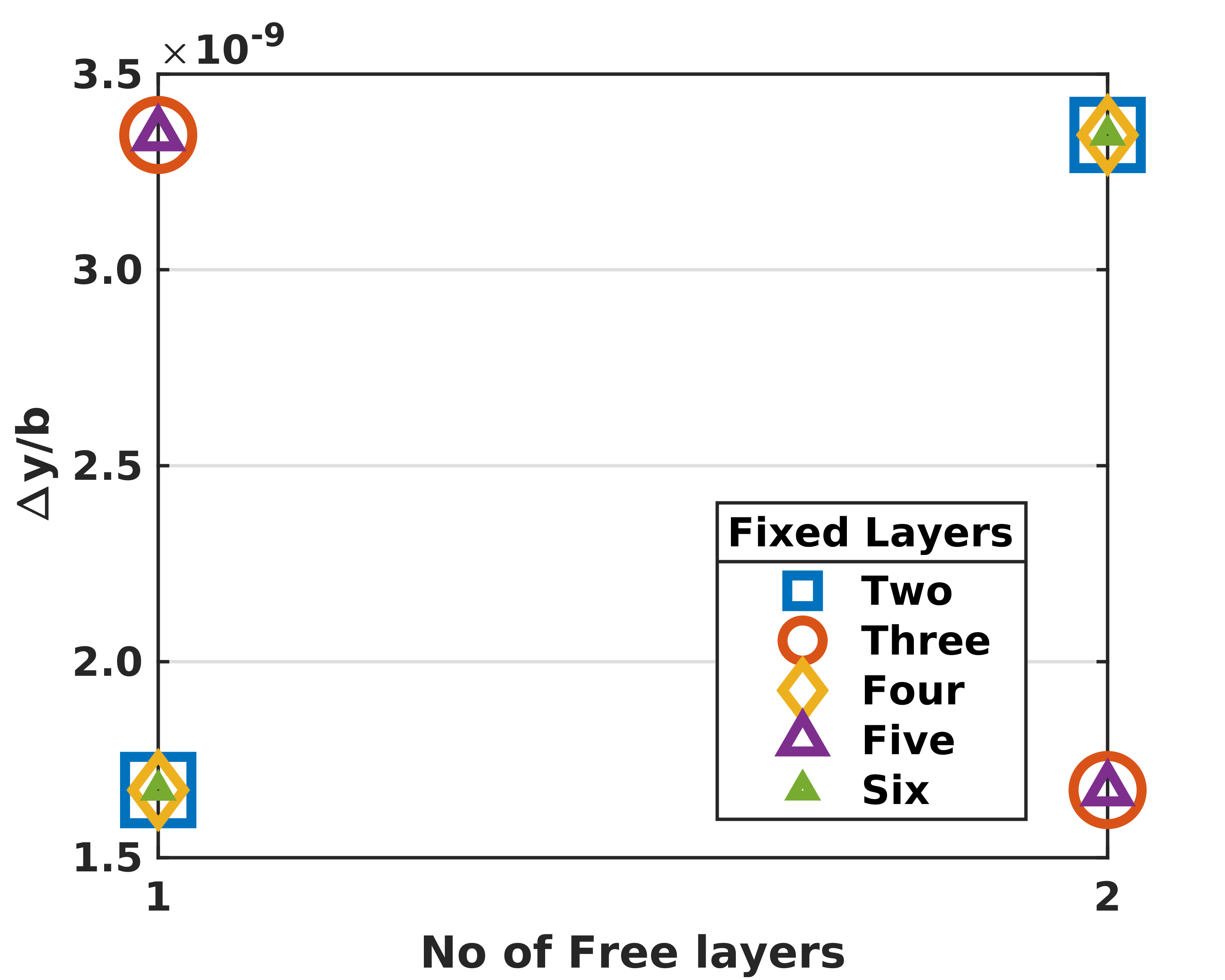}}
\qquad
\subfloat[]{
\label{}\includegraphics[height=2.18in]{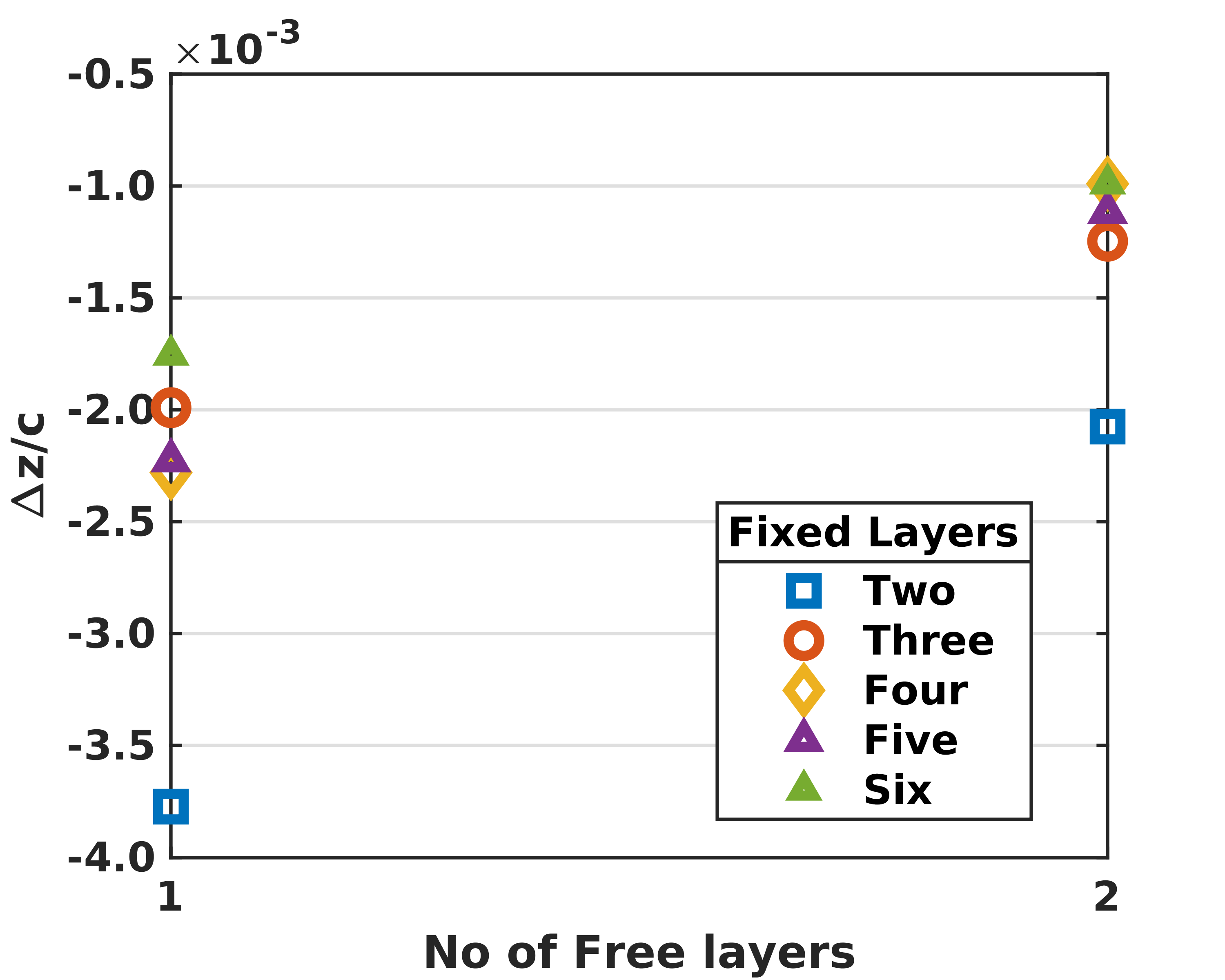}}
\qquad
\subfloat[]{
\label{surface_relaxation_no_of_fixed_layer}
\includegraphics[height=2.18in]{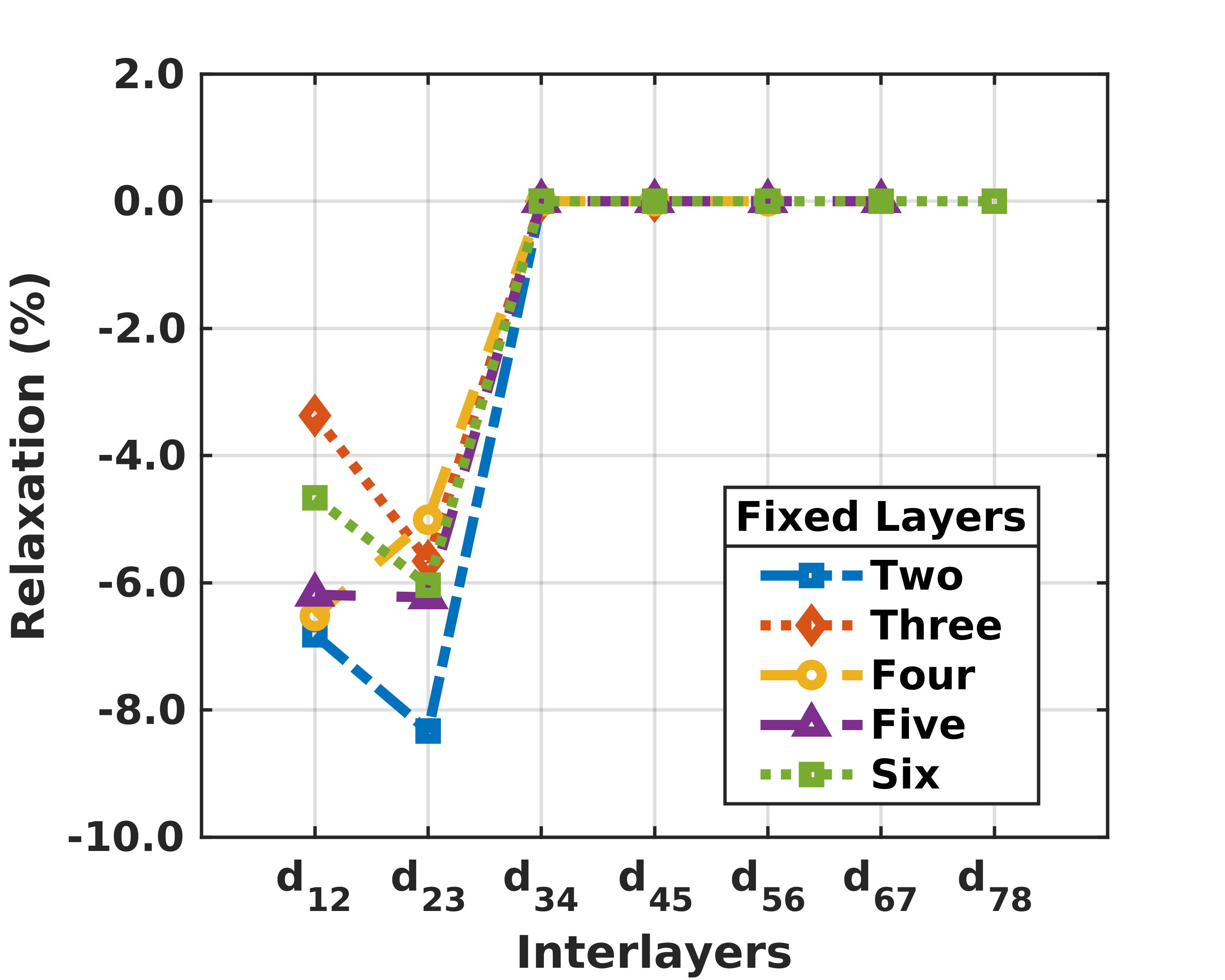}}
\caption {
The relative movement of free surface layers with respect to lattice parameters in fractional co-ordinates in (a) x, (b) y, and (c) z, and (d) Relaxation estimation, \(d_{ij}\) (i,j=1,2,3,$\ldots$) as a percentage of the bulk interlayer distance for the Zn \hkl(0001) surface. 
Similar to~\fig{no_of_free_layer_optimization}, positive and negative value specifies expansion and contraction, respectively. 
Here, we varied the number of fixed layers to determine optimal number of fixed layers fixing 2 free layers.\label{fig:no_of_fixed_layer_optimization}}
\end{figure}
\par For improving the computational efficiency it is necessary to find the minimum number of fixed layers which will be acceptable for studying the Zn surface.
\fig{no_of_fixed_layer_optimization} shows similar results to~\fig{no_of_free_layer_optimization} that surface reconstruction occurs only in the z-direction.
The degree of relaxation \(d_{ij}\) presents the shifting of the two free layers due to changing the number of fixed layers.
We find that the movement of the free layers is highest for two fixed layers as shown in~\fig{no_of_fixed_layer_optimization} (d).
All other fixed layers cases display similar second free slab movement and we choose four fixed layers for studying the Zn surface with 6-layer (2 free + 4 fixed) model. 
\begin{figure}[h]
\centering
\subfloat[]{
\label{}
\includegraphics[height=2.1in]{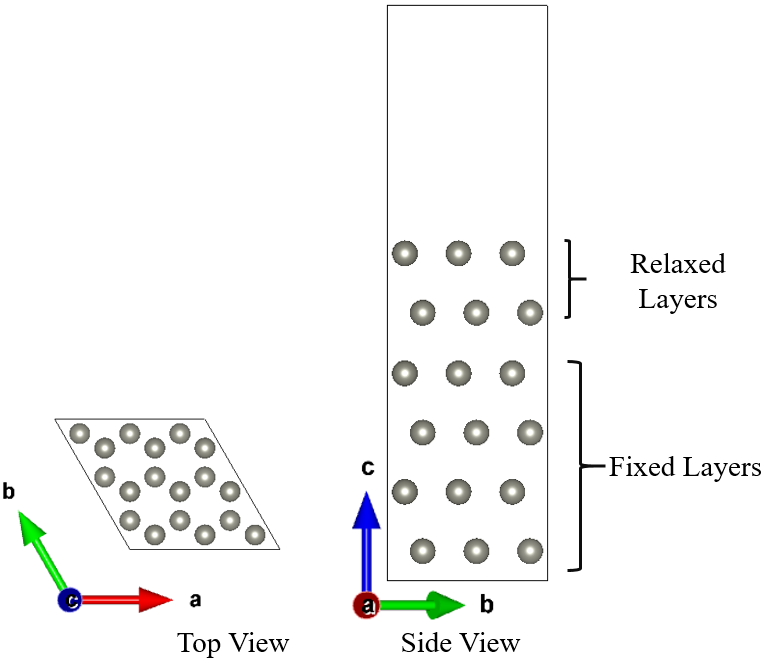}}
\qquad
\subfloat[]{
\label{}
\includegraphics[height=2.1in]{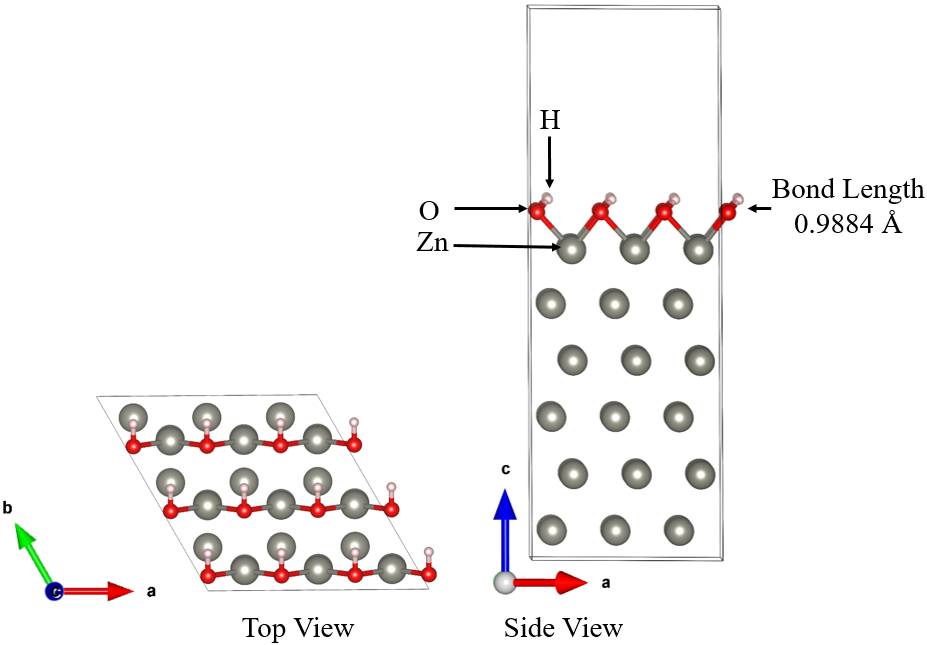}}
\qquad
\subfloat[]{
\label{}
\includegraphics[height=2.3in]{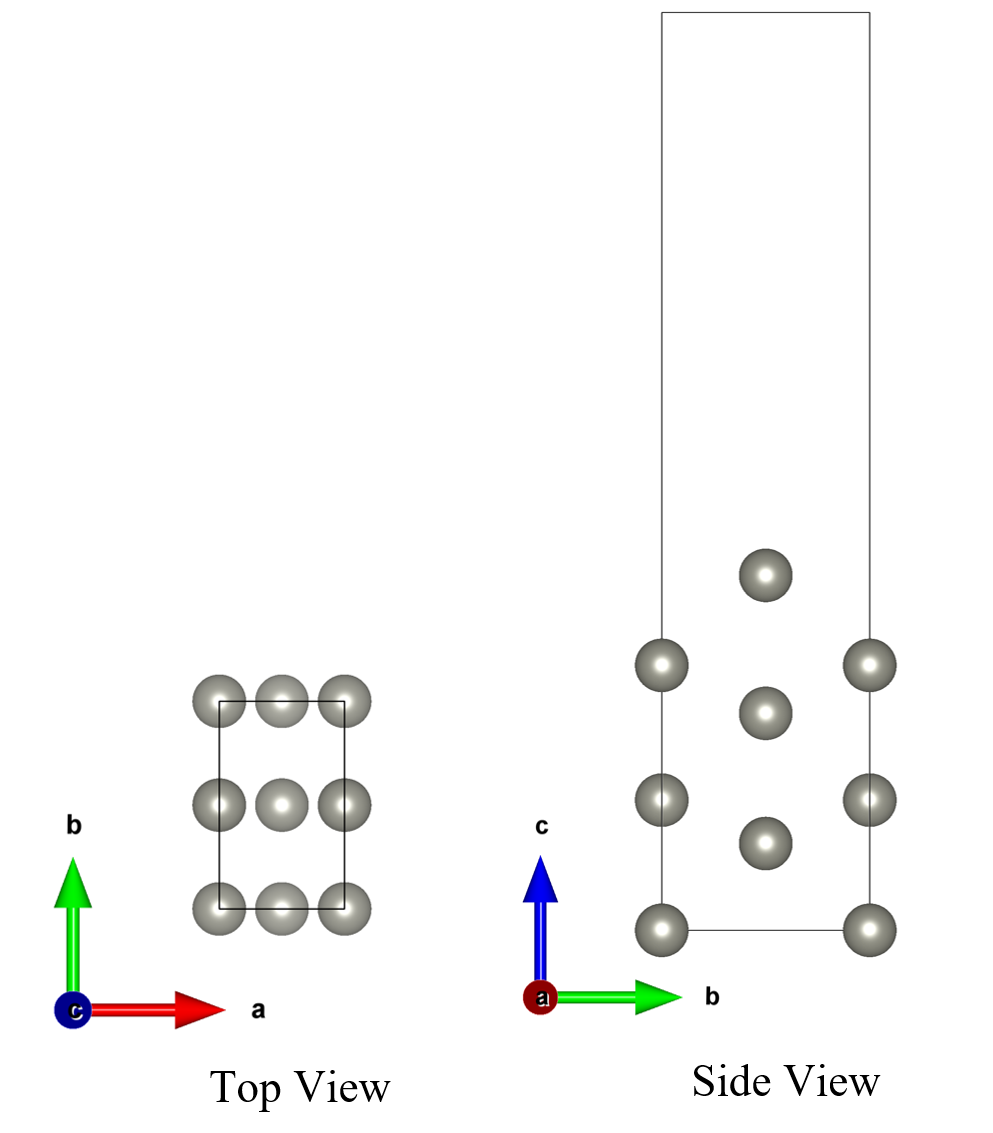}}
\qquad
\subfloat[]{
\label{}
\includegraphics[height=2.3in]{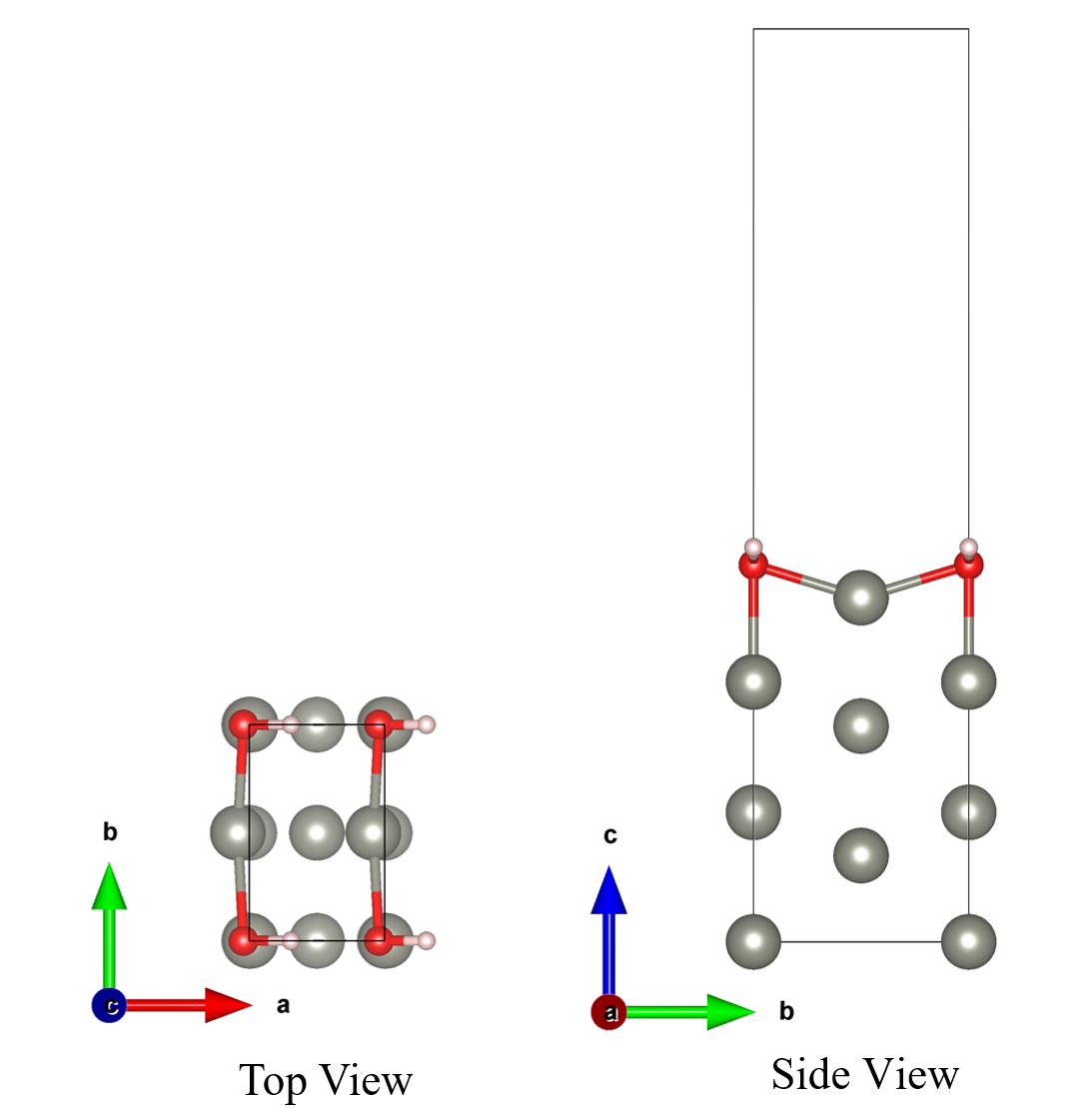}}
\qquad
\caption {
\label{fig:Zn_model}Optimized geometry of different structures (a) Zn \hkl(0001), (b) hydroxylated Zn \hkl(0001), (c) prismatic Zn \hkl(1-100), and (d) prismatic hydroxylated Zn \hkl(1-100) thin film.}
\end{figure}
\subsection{Surface Energy}
\par Surface energy (\(\gamma\)) is a basic physical criterion of metallic surface, it is essential to comprehend a wide range of surface events, such as adsorption, surface corrosion, surface segregation growth rate, {\it etc}).
As discussed in the main article, there are several approaches to evaluating surface energy to solve the problem of surface energy divergence in separate bulk and surface simulations.
\par {\it Standard approach} - The surface energy (\(\gamma\)) can be determined by the formula -
  \begin{equation} \label{eq:standard_approach}
  \begin{split} 
   \gamma(N) = \frac{1}{2.A} (E_{\textrm{slab}}(N)-N.E_{\textrm{bulk}})
   \end{split}
   \end{equation} 
Here, $\mathrm{{\it{E}}_{bulk}}$ and $\mathrm{{\it{E}}_{slab}}$ are the bulk energy per atom and slabs' total energy, respectively. 
A is the surface area, and N represents the total atoms count in the surface model. 
The denominator of~\eq{standard_approach} has a factor of 2 due to the two surfaces of the slab~\cite{Sun}.
However, it is noticed that the standard approach goes through divergence problems due to the origination of $\mathrm{{\it{E}}_{slab}}$  and $\mathrm{{\it{E}}_{bulk}}$ from two distinct sets of simulations with probable incompatibilities in the numerical calculations.
\par {\it Boettger relation}- Boettger estimated the value for $\mathrm{{\it{E}}_{bulk}}$ in~\eq{standard_approach} by~\cite{Scholz}
\begin{equation} 
   \begin{split} 
   \frac{\Delta E}{\Delta N} =  \frac{E_{\textrm{slab}}(N) - E_{\textrm{slab}}(N-1)}{N_{2}-N_{1}}  \approx E_{\textrm{bulk}}
   \end{split}
   \end{equation}
Here, $\mathrm{{\it{E}}_{bulk}}$ is calculated as the enhancement of $\mathrm{{\it{E}}_{slab}}$ by attaching one layer to the slab.
This method has the advantage of avoiding separate bulk calculations for energy value though it involves an extra computational effort.
\par {\it Linear fit method}- This method can be applied on~\eq{standard_approach} for large N which approaches convergence quickly than the previous two methods. 
  \begin{equation} 
   \begin{split} 
    E_{\textrm{slab}}(N) \approx 2. A.\gamma + N. E_{\textrm{bulk}} \\
   or, \gamma = \frac{E_{\textrm{slab}}(N) - N. E_{\textrm{bulk}}}{A}
   \end{split}
   \end{equation}
The target here is to set a straight line to the $\mathrm{{\it{E}}_{slab}}$ versus N data set and then utilize the slope of the line as the bulk energy~\cite{Sun}.
\begin{figure}[h]
\centering
\subfloat[]{
\label{Linear_fit_Standard_approach}
\includegraphics[height=2.18in]{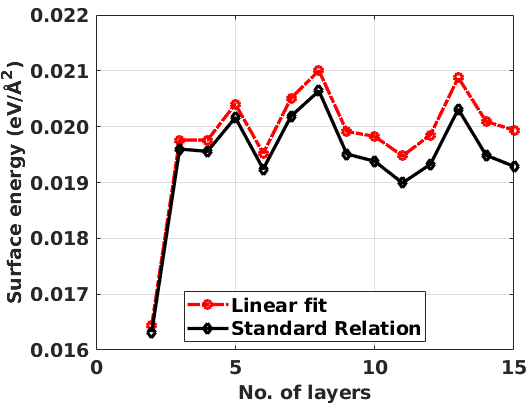}}
\qquad
\subfloat[]{
\label{Boettger_relation}
\includegraphics[height=2.18in]{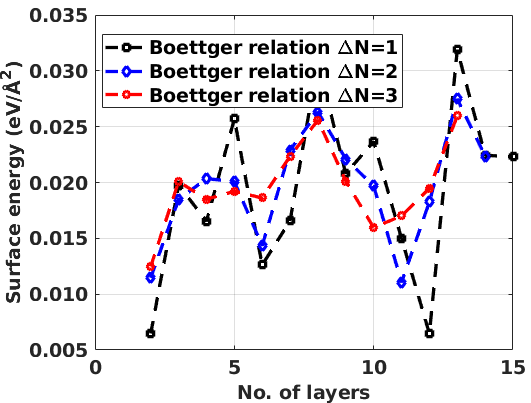}}
\caption {
Surface energies with number of layers by different methods: (a) Linear fit, and standard relation, (b) Boettger relation  in  $\mathrm{eV/\mathring{A}^2}$ as  a function of number of layers.\label{fig:surface_energy_convergence}}
\end{figure}
\par From the current case studies presented in~\fig{surface_energy_convergence}, it is inferred that the computed values of surface energy are reasonably converged with film thicknesses of 3 or more atomic layers whereas 6 atomic layers can be judged as adequately thick. 
In our study, both linear fit and standard approach show similar convergence patterns while analyzing the number of layers. 
Secondly, the Boettger method also performs fairly well in given conditions on increasing the step size.
This method was already used in the prior analysis, where so-called quantum size effects (QSEs) in the form of oscillating patterns for \(\gamma\) were found similar to~\fig{surface_energy_convergence}. 
It was noticed that increasing the step width, that is $\mathrm{{\it{N}}_{2}-{\it{N}}_{1} = 2, 3,}$ {\it etc} lessens these oscillating patterns~\cite{Scholz}.
Therefore, we also examined this option for our Zn \hkl(0001) surface and observed similar patterns. 
Furthermore, the standard approach, which applies an independent bulk energy simulation exhibits an impressive convergence and didn't diverge with increasing slab layers number.
All three methods give average surface energy of 20 $\mathrm{meV/\mathring{A}^2}$  with a deviation of less than  $\mathrm{\pm1.5\:meV/\mathring{A}^2}$ for 6 mono-crystalline layers.
Furthermore, our model's surface energy is consistent with the number of layers and with the range of previous computational studies on Zn\hkl(0001) 22 $\mathrm{meV/\mathring{A}^2}$~\cite{10.1038/sdata.2016.80}.
\subsection{Van der Waals Interactions}
\begin{figure}[H]
\centering
\includegraphics[width=0.5\textwidth]{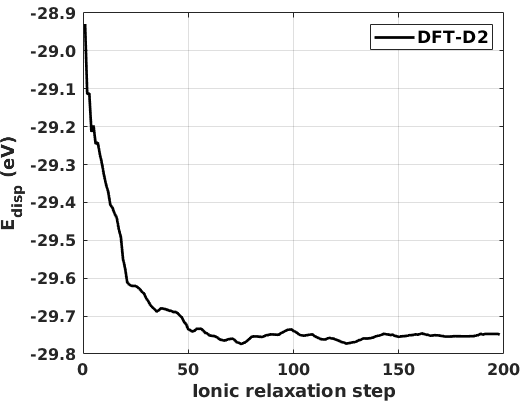}
\caption{Variation of Van der Waals interactions with ionic relaxation steps. Here, the correction term $\mathrm{E_{disp}}$ contains dispersion coefficient and damping function to evaluate interactions over a suitably chosen cutoff radius.\label{fig:vdw}}
\end{figure}
In this study, on average the effect of dispersion force is -29.6 eV as displayed in~\fig{vdw} which is nearly 5\% of total energy. 
\subsection{Bulk Elastic Properties}
\par In this study, the elastic constants for bulk Zn are determined to verify the ab initio calculations.
For a hexagonal close-packed (hcp) crystal, the five independent second-order elastic constants are \(C_{ij}: C_{11},C_{12},C_{13},C_{33},\) and \(C_{44}\) ~\cite{Vei}.
Here, the convention to denote the indices i and j in \(C_{ij}\) is xx=1, yy=2, zz=3  for the compression components, and as yz=4, zx=5, xy=6 for the shear components. 
\begin{equation} \label{eq1}
C_{ij}=\begin{pmatrix}
C_{11} & C_{12} & C_{13} & 0 & 0 & 0\\
C_{12} & C_{11} & C_{13} & 0 & 0 & 0\\
C_{12} & C_{13} & C_{33} & 0 & 0 & 0\\
0 & 0 & 0 & C_{44} & 0 & 0\\
0 & 0 & 0 & 0 & C_{44} & 0\\
0 & 0 & 0 & 0 & 0 & \frac{C_{11}-C_{12}}{2}
\end{pmatrix}
\end{equation}
\par The second-order elastic constants shown here can only characterize the linear elastic responses. 
On the other hand, to model the nonlinear elastic response of a system, we need higher $\mathrm{(>2)}$ order elastic constants~\cite{Qing}. 
By determining the eigenvalues of the stiffness matrix, sufficient and  necessary criteria for elastic stability in the hexagonal crystal are as follows-~\cite{Mouhat}
\begin{gather} 
C_{11} > |C_{12}|, \;\;2C_{13}^2 < (C_{11}+C_{12})C_{33}, \;\; C_{44}> 0, \:C_{66} > 0 
\end{gather}
\par Two closely linked approaches, namely the stress-strain method, and the energy density method (EDM), are available in determining the elastic properties from the first principles technique.
EDM computes the total energy of a crystal as a function of its pressure or volume and the second-order expansions of this energy with regard to the applied lattice strains define the elastic constants~\cite{Hongzhi}. 
After applying a small Lagrangian strain \(\epsilon\) as shown in~\tab{energy-strain approach} on unit cell, we used Taylor expansion of the strain energy density \((\Delta{E}/V_0)\) in terms of the strain tensor.
\begin{equation} 
E(V,\epsilon)=E(V_0)+V\sum_{i,j=1}^6\sigma_i\epsilon_i+\frac{V}{2}\sum_{i,j=1}^6C_{ij}\epsilon_i\epsilon_j+...
\end{equation}
\begin{equation} 
\frac{\Delta{E}}{V_0}=\frac{1}{2!}\sum_{i,j=1}^6 C_{ij}\epsilon_{i}\epsilon_{j}+O(\epsilon^3)
\end{equation}
where, \(C_{ij}\) denotes the elastic constants using Voigt notation, and \(V_{0}\) stands for the unit cell's volume ~\cite{Yufeng}.
Therefore, \(C_{ij}\) can be obtained by fitting the total energies calculated under applied strains to a parabola near the minimum energy.
The stress-strain method computes stress values triggering different strains in the crystal~\cite{Hongzhi}. 
In the present work, we used EDM to determine mechanical properties of thin film over Zn anode using ab-initio density functional theory (DFT).
\begin{table} [h]
\begin{center}
 \begin{tabular}{c c c} 
 \hline
 Strain index & Strain vector $\epsilon$ & Elastic energy \((\frac{\Delta{E}}{V})\) \\ [0.5ex]
 \hline
 1 & ($\Delta$, $\Delta$, 0, 0, 0, 0) & \((C_{11}+C_{22})\Delta^2\) \\ 
 2 & (0, 0, 0, 0, 0, $\Delta$) & \(\frac{1}{4}(C_{11}-C_{22})\Delta^2\) \\
 3 & (0, 0, $\Delta$, 0, 0, 0) & \(\frac{1}{2}C_{33}\Delta^2\) \\
 4 & (0, 0, 0, $\Delta$, $\Delta$, 0) & \(C_{44}\Delta^2\)\\
 5 & ($\Delta$, $\Delta$, $\Delta$, 0, 0, 0) & \((C_{11}+C_{22}+2C_{13}+\frac{C_{33}}{2})\Delta^2\) \\ [1ex] 
 \hline
\end{tabular}
\caption{\label{tab:energy-strain approach} List  of  applied strain  modes  to  computed  elastic  constants  based  on  energy-strain approach for hexagonal system.~\cite{Vei}}
\end{center}
\end{table}

The Voigt-Reuss-Hill (VRH) approximations can be used for anisotropic single-crystal system to estimate the mechanical properties such as shear (G), bulk (B), Young's modulus (E) and the Pugh ratio \((\gamma)\) in terms of an isotropic polycrystalline system~\cite{Hill, Kyoungmin, 10.1063/1.1709944}.
Firstly, using Voigt approximation, the upper limit (Voigt) of G and B are computed from \((C_{ij} )\) as 
\begin{gather} 
B_{\textrm{V}} = \frac{(C_{11}+C_{22}+C_{33})+2(C_{12}+C_{23}+C_{31})}{9} \\
G_{\textrm{V}} = \frac{(C_{11}+C_{22}+C_{33})-(C_{12}+C_{23}+C_{31})+3(C_{44}+C_{55}+C_{66})}{9} 
\end{gather}
Secondly, using Reuss approximation, the lower limit of B and G are computed as,
\begin{gather} 
B_{\textrm{R}}=\frac{1}{(S_{11}+S_{22}+S_{33})+2(S_{12}+S_{23}+S_{31})} \\
G_{\textrm{R}}=\frac{15}{4(S_{11}+S_{22}+S_{33})-4(S_{12}+S_{23}+S_{31})+3(C_{44}+C_{55}+C_{66})}
\end{gather}
where the compliance tensor matrix, \(S_{ij}\) is specified as $ S_{ij} = C_{ij}^{-1}$.
Finally, B and G can be expressed using the Hill approximation as the average of Voigt and Reuss limit as,
\begin{gather} 
B =\frac{B_{\textrm{V}}+B_{\textrm{R}}}{2} \;\;and\;\; G =\frac{G_{\textrm{V}}+G_{\textrm{R}}}{2}
\end{gather} 
Using the calculated B and G from VRH approximations, Young's modulus (E), Poisson's ratio \((\nu)\),  and Pugh's ratio \((k)\) are defined as,
\begin{gather}
E = \frac{9GB}{(G+3B)}, \:\: \nu = \frac{3B-2G}{2(G+3B)},  \:\: k = \frac{B}{G}
\end{gather} 
\par From~\tab{zinc_crystal_elastic_constants}, we computed the elastic properties of bulk Zn and compared them with experimental studies. 
The benchmark of ductility, namely Poisson's ratio, and Pugh's ratio, indicates that bulk Zn is below the ductility cutoff which is  0.26 for $\nu$ and 1.75 for k~\cite{Michael}. 
\begin{table}[h]
\small
\setlength\tabcolsep{4pt}
 \begin{center}
\begin{tabular}{c c c c c c c c c c c c} 
\hline
 & $C_{11}$ & $C_{33}$ & $C_{44}$ & $C_{66}$ & $C_{12}$ & $C_{13}$ & B & G & E & $\nu$ & $k$\\ [0.5ex]
 \hline
Present study & 178.1 & 47.3 & 30.6 & 67.5 & 43.2 & 41.2 & 59.8 & 38.1 & 94.3 & 0.24 & 1.57 \\ [1ex] 
Garland et al. (0 K)~\cite{10.1103/PhysRev.111.1232} & 177.0 & 68.5 & 45.9 & 71.1 & 34.8 & 52.8 & - & - & - & - & - \\ [1ex] 
H. M. Ledbetter \\ (room temperature)~\cite{10.1063/1.555564} & 163.0 & 60.3 & 39.4 & 65.9 & 30.6 & 48.1 & 68.3 & 39.9 & 100.3 & 0.26 & 1.71\\ [1ex]
 \hline
\end{tabular}
 \caption{\label{tab:zinc_crystal_elastic_constants} List  of second-order elastic-stiffness coefficients $(C_{ij})$ (GPa), Bulk modulus (B) (GPa), Shear modulus (G) (GPa), Young's modulus (E) (GPa), Poisson's ratio \((\nu)\) and Pugh ratio (k) of zinc.}
\end{center}
\end{table}

\subsection{Energy Profile of Strained Surfaces}
\begin{figure}[H]
\centering
\subfloat[]{
\label{}
\includegraphics[height=2.25in]{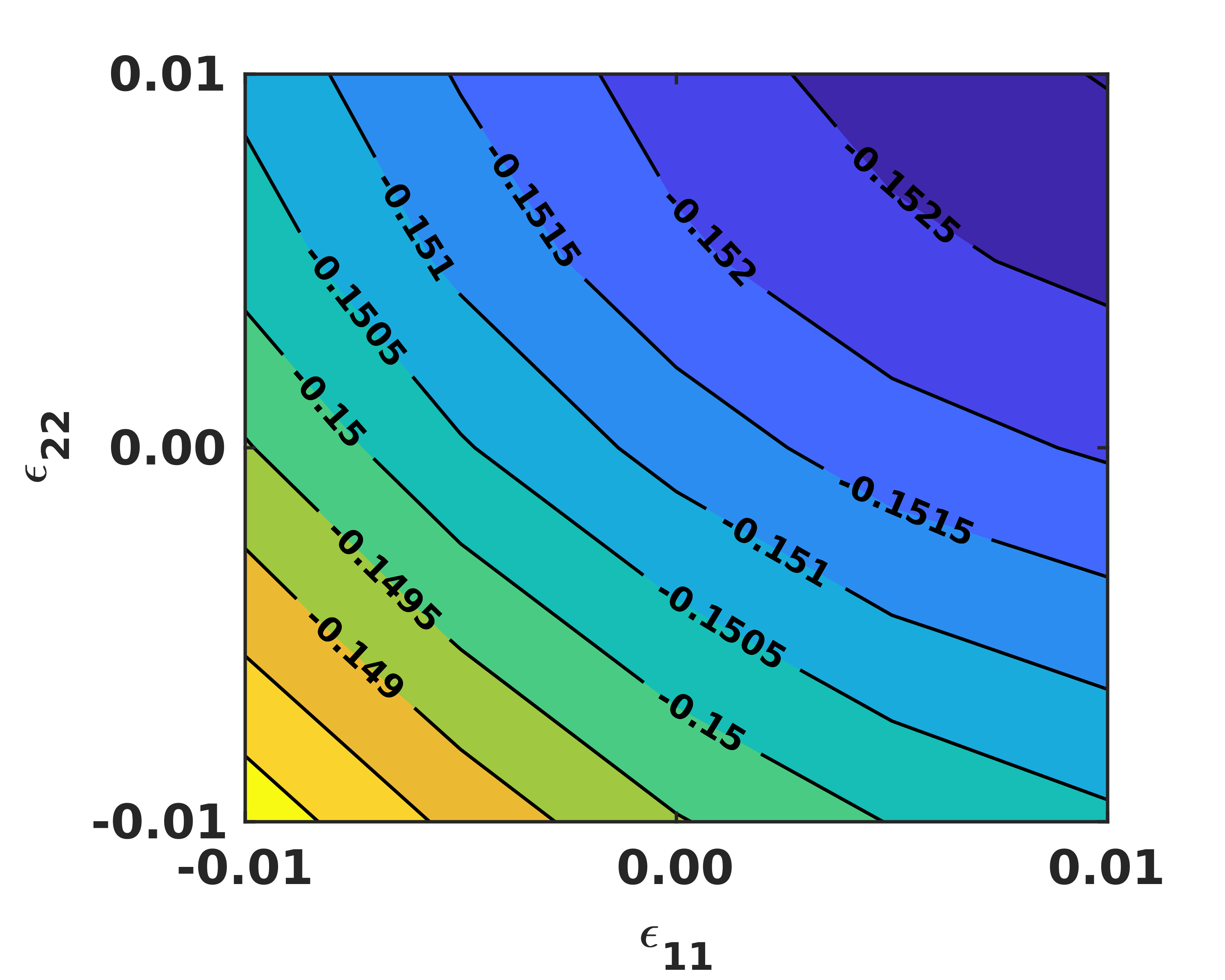}}
\qquad
\subfloat[]{
\label{}
\includegraphics[height=2.25in]{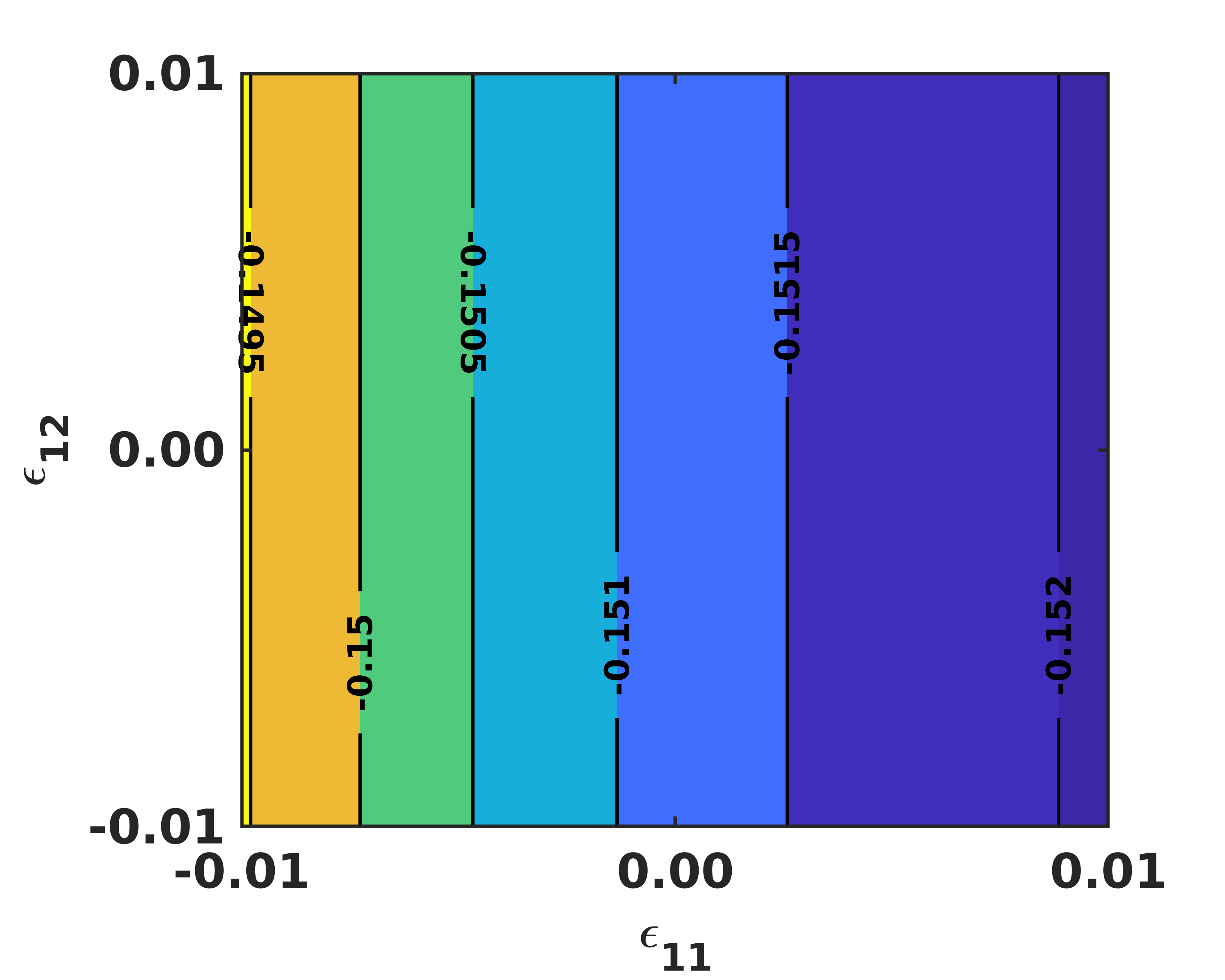}}
\qquad
\subfloat[]{
\label{}
\includegraphics[height=2.25in]{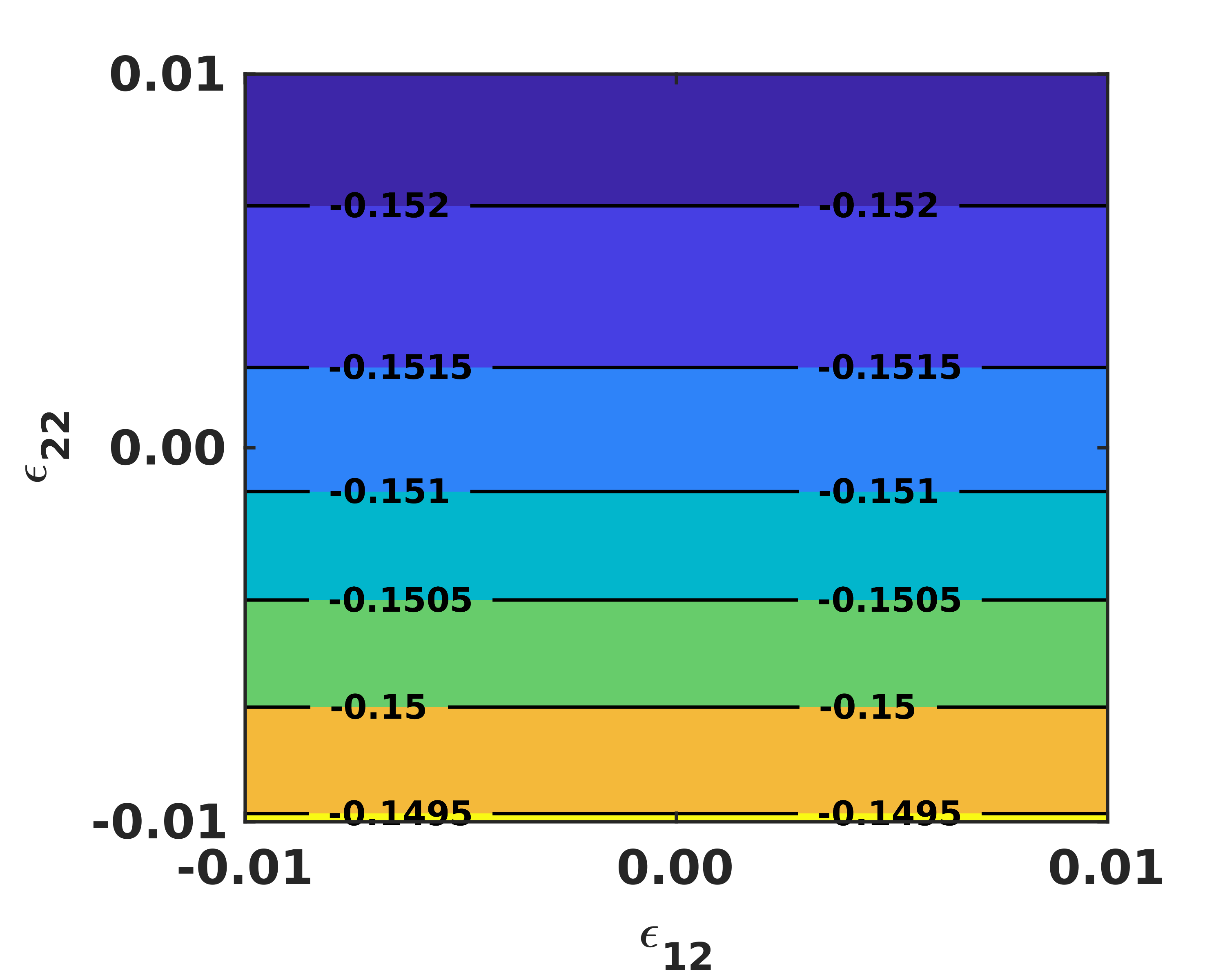}}
\qquad
\subfloat[]{
\label{}
\includegraphics[height=2.25in]{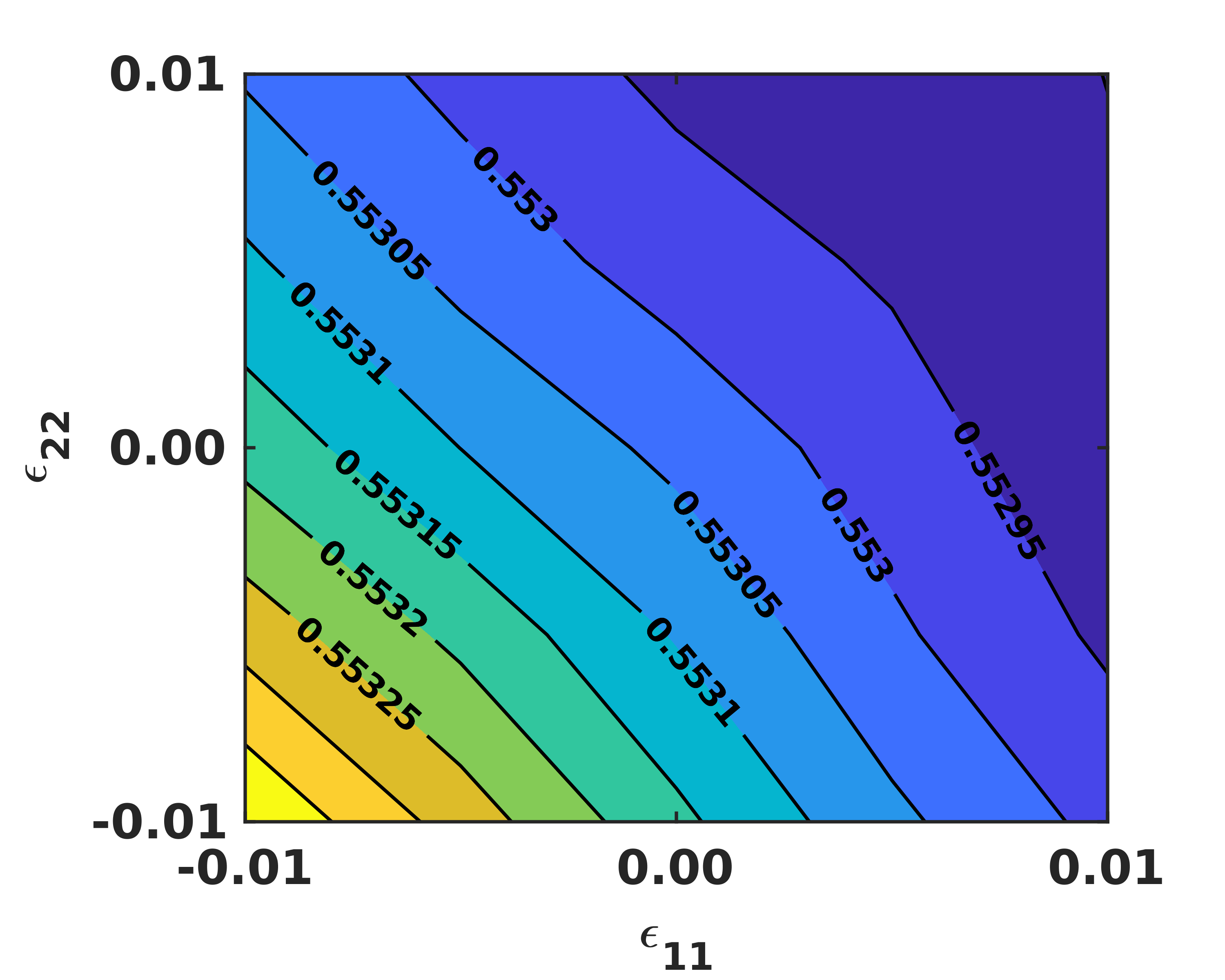}}
\qquad
\caption {\label{fig:surface_energy_with_strain_bulk_Zn_ZNOH}Hydroxylated Zn \hkl(0001) thin film: Contour plot of variation of surface energy ($\mathrm{eV/\textrm{\r{A}}^{2}}$) with biaxial strain for (a) 11-22, (b) 11-12, (c) 12-22 direction, and (d) Zn \hkl(0001) thin film: Contour plot of variation of surface energy ($\mathrm{eV/\textrm{\r{A}}^{2}}$) with biaxial strain for 11-22 direction.}
\end{figure}
\begin{figure}[H]
\centering
\subfloat[]{
\label{}
\includegraphics[height=2.25in]{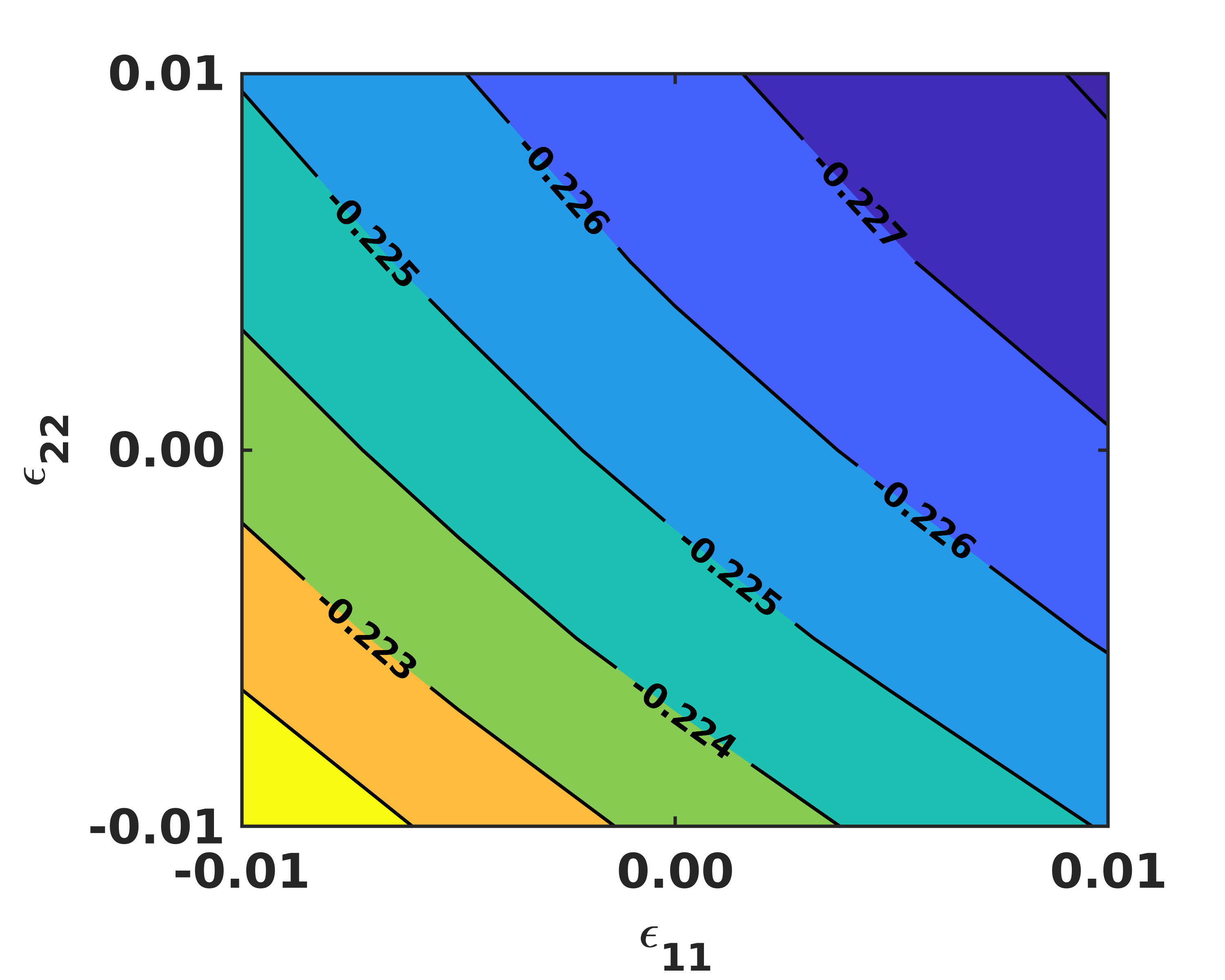}}
\qquad
\subfloat[]{
\label{}
\includegraphics[height=2.25in]{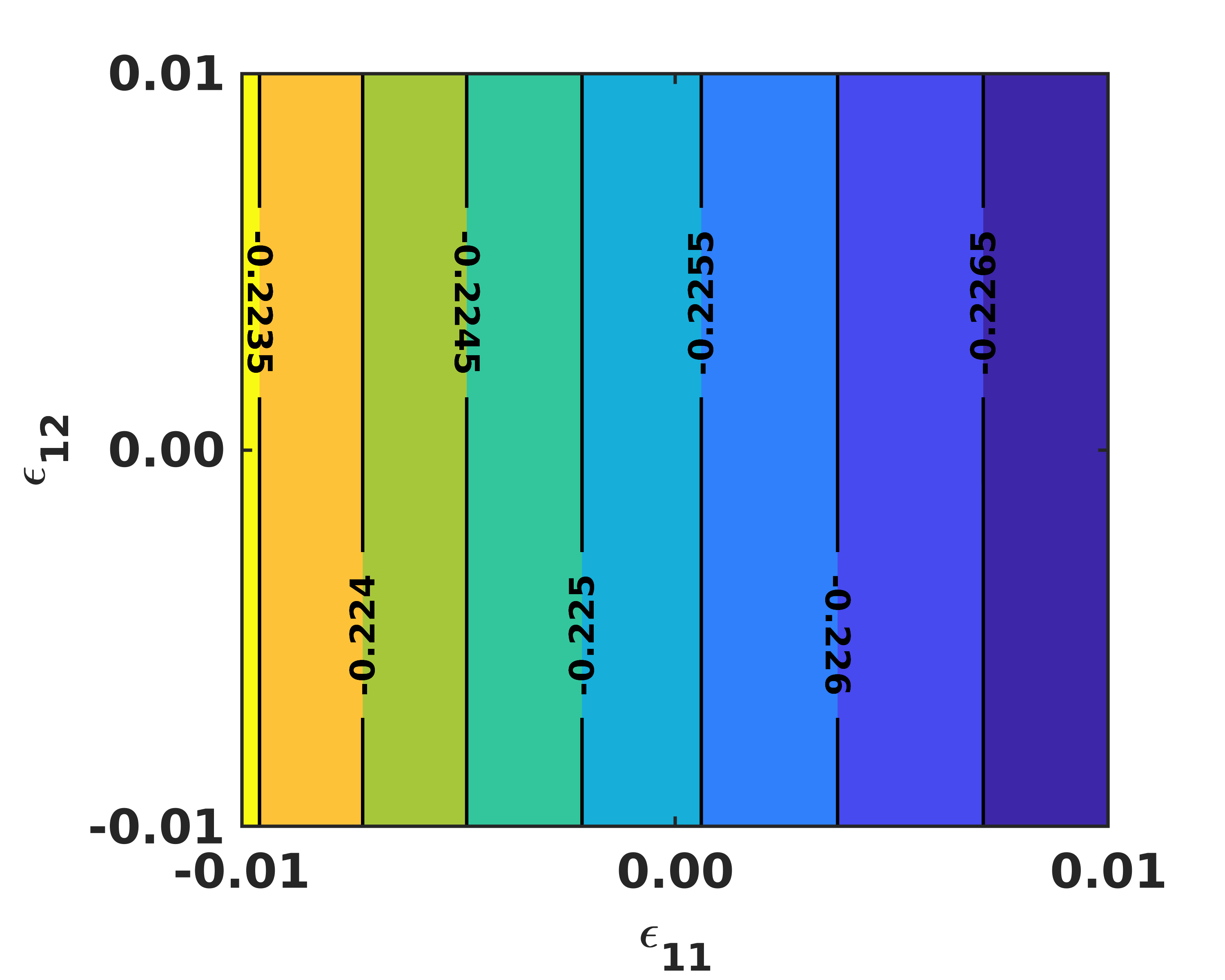}}
\qquad
\subfloat[]{
\label{}
\includegraphics[height=2.25in]{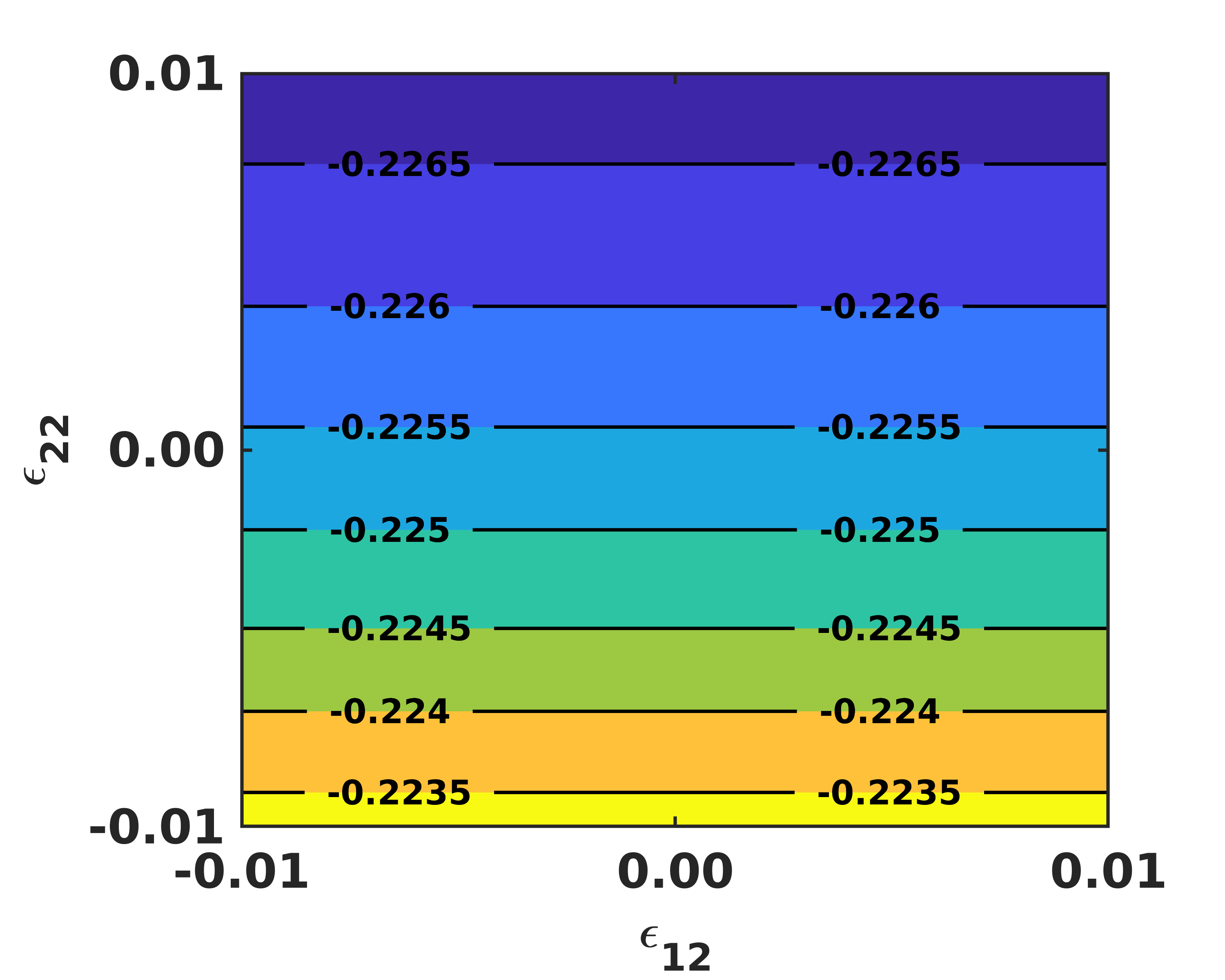}}
\qquad
\caption {\label{fig:surface_energy_with_strain_prismatic_alucone}Contour plot of variation of surface energy ($\mathrm{eV/\textrm{\r{A}}^{2}}$) with biaxial strain for alucone coated hydroxylated Zn \hkl(1-100) surface (a) 11-22, (b) 11-12, and (c) 12-22 direction.}
\end{figure}

\subsection{Interpolating the Strain-Dependent Surface Energy}
\begin{table}[h]
\small
\setlength\tabcolsep{4pt}
 \begin{center}
\begin{tabular}{c c c c c } 
\hline
Zinc \hkl(0001) surface & FDM (4th order) & LME & HOLMES \\ [0.5ex]
 \hline
$d_{11}\:({d{\gamma}/d{\epsilon}_{11}}$) & -0.1 & -0.1 & -0.1 \\ [1ex]
$d_{22}\:({d{\gamma}/d{\epsilon}_{22}}$) &-0.1 & -0.1 & -0.1  \\ [1ex]
$d^{'}_{11}\:({d^{2}{\gamma}/d{\epsilon}_{11}}^{2}$) & -1.6  & -0.5  & -1.4 \\ [1ex] 
$d^{'}_{22}\:({d^{2}{\gamma}/d{\epsilon}_{22}}^{2}$) & -1.4 & -0.3  & -1.4  \\ [1ex] 
$d^{'}_{12}\:({d^{2}{\gamma}/d{\epsilon}_{11}}d{\epsilon}_{22}$) & 11.7 & 9.6 & 11.5 \\ [1ex] 
 \hline
Hydroxylated Zinc \hkl(0001) surface & FDM (4th order) & LME & HOLMES \\ [0.5ex]
 \hline
$d_{11}\:({d{\gamma}/d{\epsilon}_{11}}$) & -0.1 & -0.1 & -0.1  \\ [1ex]
$d_{22}\:({d{\gamma}/d{\epsilon}_{22}}$) & -0.2 & -0.2 &  -0.2 \\ [1ex]
$d^{'}_{11}\:({d^{2}{\gamma}/d{\epsilon}_{11}}^{2}$) & 6.4  & 7.0  & 6.6 \\ [1ex] 
$d^{'}_{22}\:({d^{2}{\gamma}/d{\epsilon}_{22}}^{2}$) & 6.9 & 6.7  & 6.7 \\ [1ex] 
$d^{'}_{12}\:({d^{2}{\gamma}/d{\epsilon}_{11}}d{\epsilon}_{22}$) & 0.7 & 0.6 & 0.6 \\ [1ex] 
 \hline
Alucone deposited \\
Hydroxylated Zinc \hkl(0001) surface & FDM (4th order) & LME & HOLMES \\ [0.5ex]
 \hline
$d_{11}\:({d{\gamma}/d{\epsilon}_{11}}$) & -0.2 & -0.2 &  -0.2 \\ [1ex]
$d_{22}\:({d{\gamma}/d{\epsilon}_{22}}$) & -0.2 & -0.2 & -0.2  \\ [1ex]
$d^{'}_{11}\:({d^{2}{\gamma}/d{\epsilon}_{11}}^{2}$) & 1.8 & 2.5 & 2.3  \\ [1ex] 
$d^{'}_{22}\:({d^{2}{\gamma}/d{\epsilon}_{22}}^{2}$) & 8.9 & 8.5  & 8.8  \\ [1ex] 
$d^{'}_{12}\:({d^{2}{\gamma}/d{\epsilon}_{11}}d{\epsilon}_{22}$) & 3.6 & 3.5 & 2.3 \\ [1ex] 
 \hline
\end{tabular}
 \caption{\label{tab:FDM_LME_HOLMES} Comparison among Finite difference method (FDM), Local Maximum Entropy(LMS) and Higher Order LME Scheme (HOLMES) method for slope ($\mathrm{d}$) and curvature ($\mathrm{d^{'}}$) to evaluate surface elastic constants ($C_{ijkl}$).}
\end{center}
\end{table}
\begin{table}[h]
\small
\setlength\tabcolsep{4pt}
 \begin{center}
\begin{tabular}{c c c c c } 
\hline
Zinc \hkl(1-100) surface & FDM (4th order) & LME & HOLMES \\ [0.5ex]
 \hline
$d_{11}\:({d{\gamma}/d{\epsilon}_{11}}$) & -0.01 & -0.01 & -0.01 \\ [1ex]
$d_{22}\:({d{\gamma}/d{\epsilon}_{22}}$) & -0.01 & -0.01 & -0.01 \\ [1ex]
$d^{'}_{11}\:({d^{2}{\gamma}/d{\epsilon}_{11}}^{2}$) & 6.2 & 6.2 & 6.2\\ [1ex] 
$d^{'}_{22}\:({d^{2}{\gamma}/d{\epsilon}_{22}}^{2}$) & 3.9 & 3.6 & 3.8\\ [1ex] 
$d^{'}_{12}\:({d^{2}{\gamma}/d{\epsilon}_{11}}d{\epsilon}_{22}$) & 0.5 & 0.7 & 0.5\\ [1ex] 
 \hline
Hydroxylated Zinc \hkl(1-100) surface & FDM (4th order) & LME & HOLMES \\ [0.5ex]
 \hline
$d_{11}\:({d{\gamma}/d{\epsilon}_{11}}$) & -0.2 & -0.2 &  -0.2\\ [1ex]
$d_{22}\:({d{\gamma}/d{\epsilon}_{22}}$) & -0.2 & -0.2 & -0.2\\ [1ex]
$d^{'}_{11}\:({d^{2}{\gamma}/d{\epsilon}_{11}}^{2}$) & 6.1 & 5.8 & 4.9\\ [1ex] 
$d^{'}_{22}\:({d^{2}{\gamma}/d{\epsilon}_{22}}^{2}$) & 6.9 & 7.0 & 7.3\\ [1ex] 
$d^{'}_{12}\:({d^{2}{\gamma}/d{\epsilon}_{11}}d{\epsilon}_{22}$) & 1.9 & 2.2 & 2.2\\ [1ex] 
 \hline
Alucone deposited \\
Hydroxylated Zinc \hkl(1-100) surface & FDM (4th order) & LME & HOLMES \\ [0.5ex]
 \hline
$d_{11}\:({d{\gamma}/d{\epsilon}_{11}}$) & -0.2 & -0.2 & -0.2\\ [1ex]
$d_{22}\:({d{\gamma}/d{\epsilon}_{22}}$) & -0.2 & -0.2 & -0.2\\ [1ex]
$d^{'}_{11}\:({d^{2}{\gamma}/d{\epsilon}_{11}}^{2}$) & 6.1 & 5.8 & 5.9\\ [1ex] 
$d^{'}_{22}\:({d^{2}{\gamma}/d{\epsilon}_{22}}^{2}$) & 6.9 & 7.0 & 6.9 \\ [1ex] 
$d^{'}_{12}\:({d^{2}{\gamma}/d{\epsilon}_{11}}d{\epsilon}_{22}$) & 1.9 & 2.2 & 1.9\\ [1ex] 
 \hline
\end{tabular}
\end{center}
\end{table}

In this study, we used the Finite difference method (FDM)~\cite{10.1016/S0377-0427(00)00507-0}, Local Maximum Entropy (LME)~\cite{10.1002/nme.1534} and the Higher Order LME Scheme (HOLMES) ~\cite{10.1007/978-3-642-32979-1_7}, to calculate 1st and 2nd order derivatives from strain-energy data. 
LME and HOLMES are both mesh-free interpolation schemes designed to work over unstructured data, and whose shape function's width is governed by a parameter $\xi$ which corresponds to the rate of Gaussian decay.
From~\tab{FDM_LME_HOLMES}, all three approaches provide data in a similar range and we used derivatives from HOLMES approach to evaluate elastic properties.
For Finite Differences, $\mathrm{O(h^{2}})$ and $\mathrm{O(h^{4}})$  central difference schemes are performed in 2D for mixed derivatives whereas arbitrary order central differences are used in one dimension. 
In general, the agreement between different orders of accuracy deviates due to approximation errors in data. LME, which is a smoothing interpolation technique, provides results that seem more robust given the approximation error. 
Locality parameters ($\xi$) of 1.8 and 4.0 were tested.
In general, the larger value catches local features better, but results are reasonable either way. 
This is an $\mathrm{O(h^{2}})$ method for Hessian which is not applicable near the boundary of the domain. 
HOLMES is a higher-order smoothing interpolation technique where $\xi$ of 0.8 to 4.0 has been tested. 
HOLMES delivers the best results compared to the other two methods as it used an $\mathrm{O(h^{2}})$ hessian approximation to avoid over-fitting with smoothing interpolation to compromise robustness and higher order accuracy.
However, it is not applicable near the boundary of the domain.

\subsection{Size Dependency of Elastic Properties}
\begin{figure}[h]
\centering
\subfloat[]{
\label{}
\includegraphics[height=2.4in]{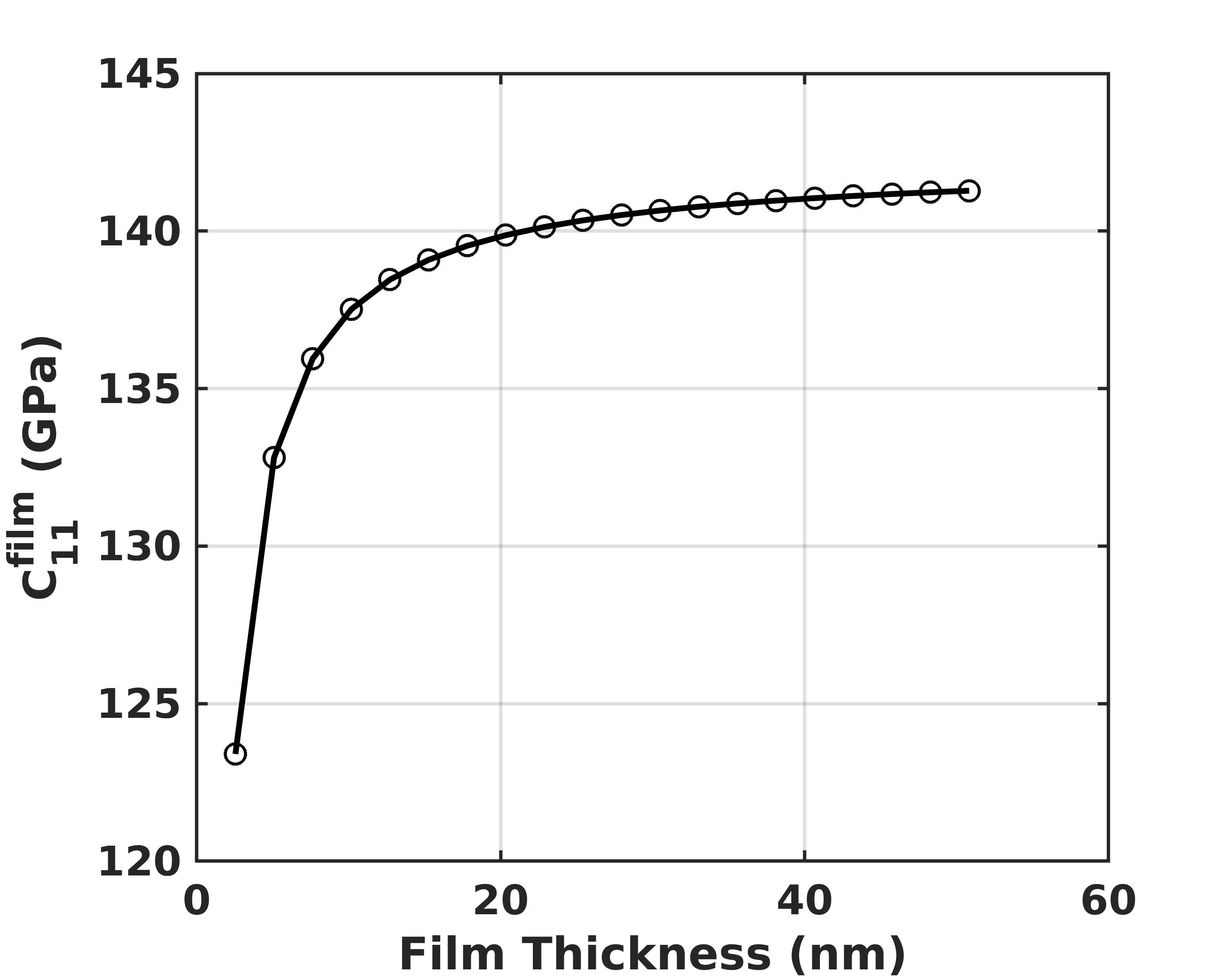}}
\qquad
\subfloat[]{
\label{}
\includegraphics[height=2.4in]{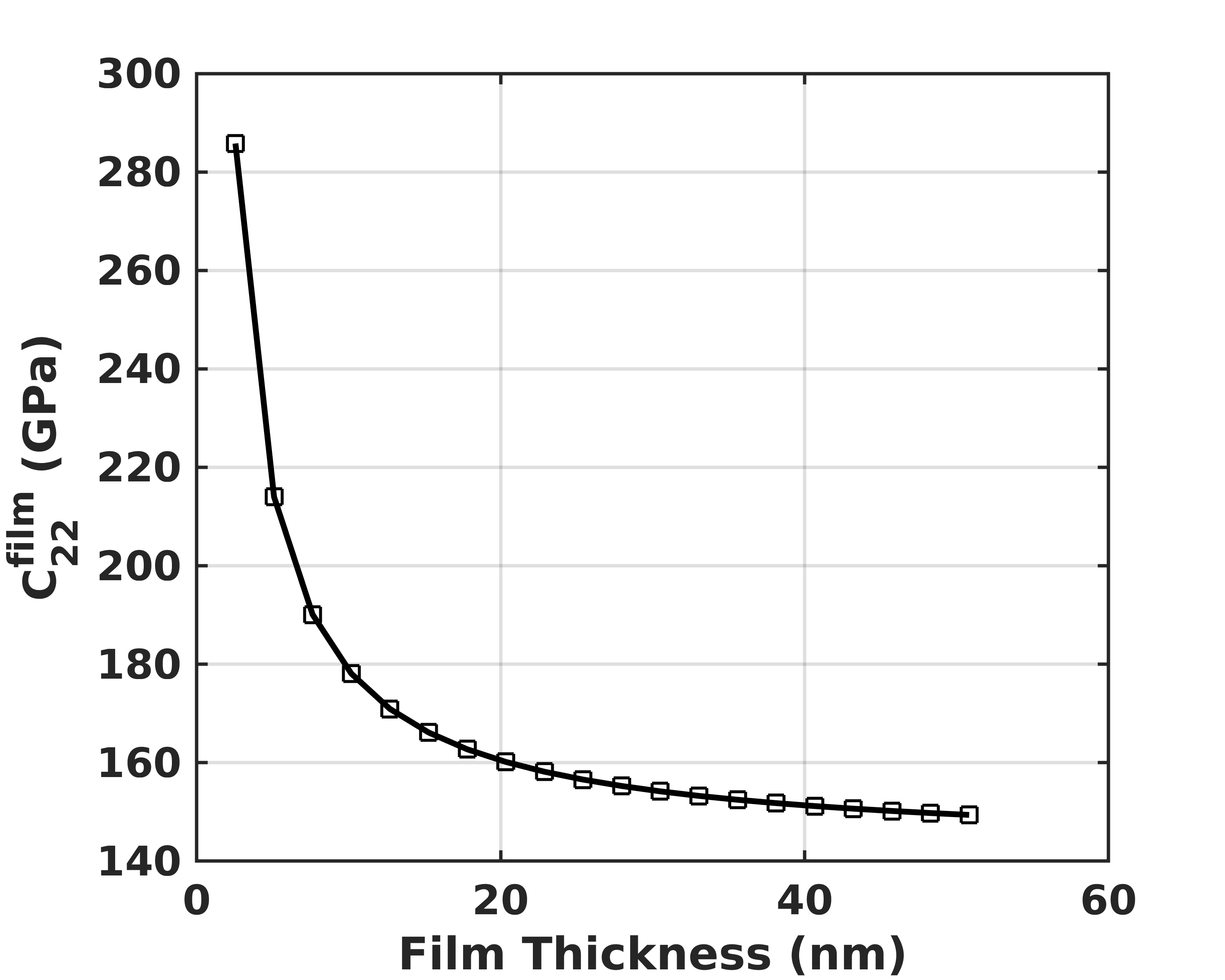}}
\qquad
\subfloat[]{
\label{}
\includegraphics[height=2.4in]{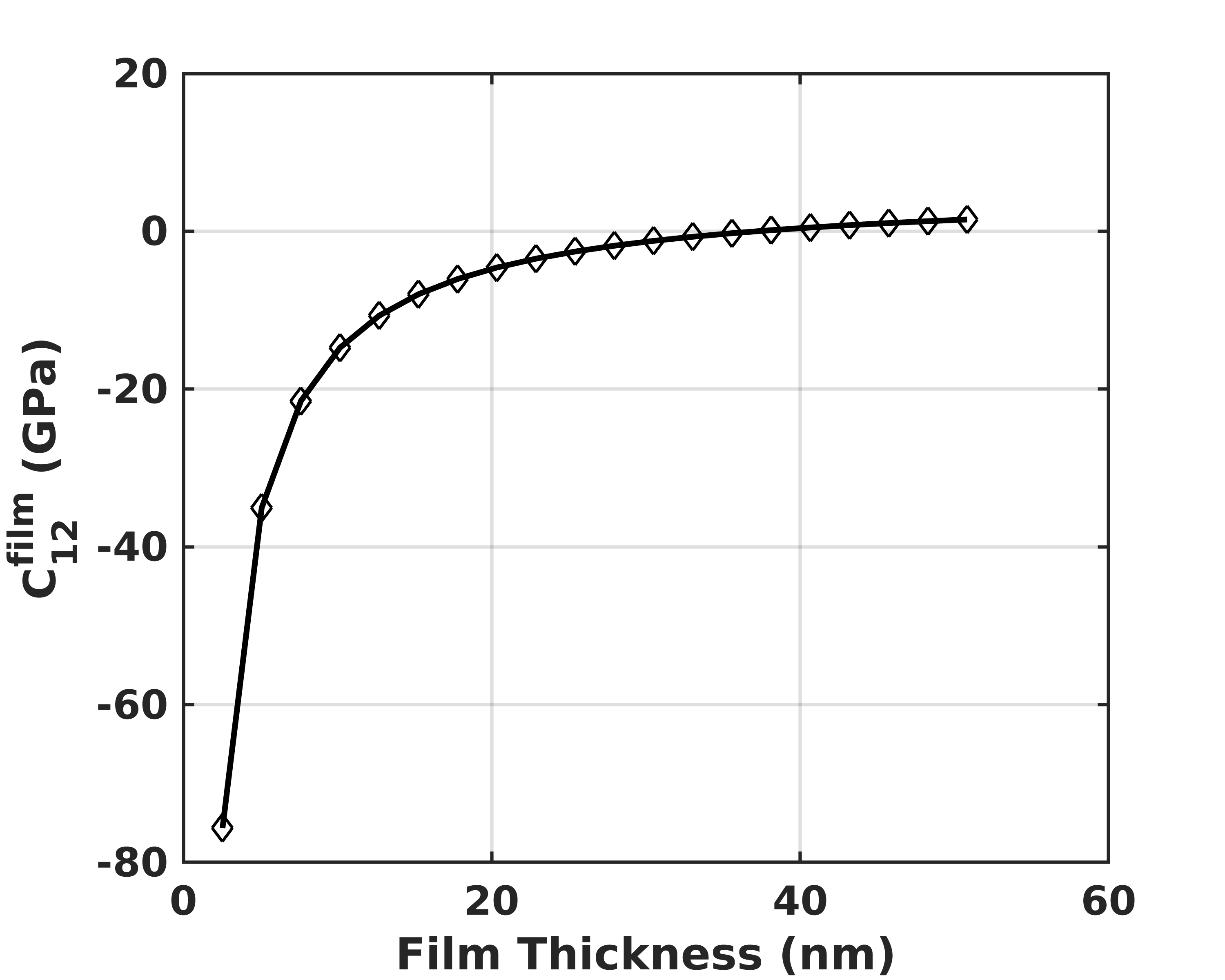}}
\qquad
\caption {\label{fig:size_dependent_elastic_properties} Variation of elastic modulus tensors with thickness for Zn (0001) thin film (a) $\mathrm{C_{11}^{film}}$, (b) $\mathrm{C_{22}^{film}}$ and (c) $\mathrm{C_{12}^{film}}$.}
\end{figure}
It is seen from~\fig{size_dependent_elastic_properties} that the computed Zn thin film modulus differs in a definite way when plotted with film thickness. 
It is because of the size dependency of elastic properties of nano-structured thin films. 
The existence of the free surfaces in a thin film is the underlying cause for the above responses~\cite{10.1088/0957-4484/11/3/301}. 
As the thin film is stretched or compressed, strain energy is accumulated both on the surface and in the bulk. 
In the context of an equilibrium viewpoint, both bulk stresses and surface stresses contribute to the force across the thin film's cross-section. 
As the effects of surface stress increase significantly with the decreasing thickness of the plate, the thin film elastic constants change exponentially with the decreasing thickness.

\providecommand{\latin}[1]{#1}
\providecommand*\mcitethebibliography{\thebibliography}
\csname @ifundefined\endcsname{endmcitethebibliography}
  {\let\endmcitethebibliography\endthebibliography}{}

\end{document}